\UseRawInputEncoding
\documentclass[aps,showpacs,prd,twocolumn,nofootinbib,nobibnotes,]{revtex4-1}
\usepackage{graphicx}
\usepackage{amssymb}
\usepackage{amsmath}
\usepackage{color}
\usepackage{float}
\usepackage{ulem}
\usepackage{accents}
\usepackage{graphicx}
\usepackage{graphicx}
\usepackage{amsfonts}
\usepackage[colorlinks=true,
pdfstartview=FitV,linkcolor=blue,
citecolor=blue,urlcolor=blue,breaklinks=true]
{hyperref}
\usepackage{array}
\usepackage{float}
\usepackage{placeins}
\usepackage[dvipsnames]{xcolor}
\usepackage{csquotes}
\usepackage{bbold}
\usepackage{units}
\usepackage{dsfont}
\usepackage{upgreek}

\newcolumntype{C}[1]{>{\centering\arraybackslash}m{#1}}

\renewcommand{\eqref}[1]{\mbox{Eq.~(\ref{#1})}}

\definecolor{ForestGreen}{rgb}{0.13,0.55,0.13}

\newcommand{\orcid}[1]{\href{https://orcid.org/#1}{\includegraphics[width=10pt]{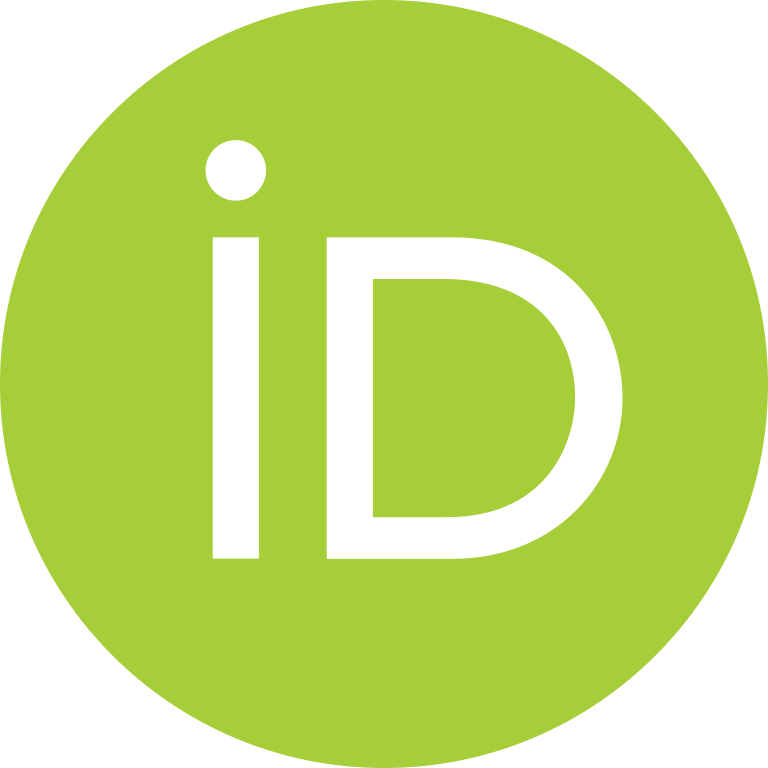}}}

\usepackage{fixmath}

\begin{document}

\title{Effects of \textit{CPT}-odd terms of dimensions three and five on electromagnetic propagation in continuous matter}

\author{Pedro D. S. Silva\orcid{0000-0001-6215-8186}$^a$}
\email{pedro.dss@discente.ufma.br}
\author{Let\'icia Lisboa-Santos\orcid{0000-0003-4939-3856}$^a$}
\email{leticia.lisboa@discente.ufma.br}
\affiliation{$^a$Departamento de F\'{i}sica, Universidade Federal do Maranh\~{a}o, Campus
Universit\'{a}rio do Bacanga, S\~{a}o Lu\'is (MA), 65080-805, Brazil}
\author{Manoel M. Ferreira Jr.\orcid{0000-0002-4691-8090}$^a$}
\email{manoel.messias@ufma.br, manojr.ufma@gmail.com}
\affiliation{$^a$Departamento de F\'{i}sica, Universidade Federal do Maranh\~{a}o, Campus
Universit\'{a}rio do Bacanga, S\~{a}o Lu\'is (MA), 65080-805, Brazil}
\author{Marco Schreck\orcid{0000-0001-6585-4144}$^a$}
\email{marco.schreck@ufma.br}
\affiliation{$^a$Departamento de F\'{i}sica, Universidade Federal do Maranh\~{a}o, Campus
	Universit\'{a}rio do Bacanga, S\~{a}o Lu\'is (MA), 65080-805, Brazil}

\begin{abstract}

In this work we study how \textit{CPT}-odd Maxwell-Carroll-Field-Jackiw (MCFJ) electrodynamics as well as a dimension-5 extension of it affect the optical activity of continuous media. The starting point is dimension-3 MCFJ electrodynamics in matter whose modified Maxwell equations, permittivity tensor, and dispersion relations are recapitulated. Corresponding refractive indices are achieved in terms of the frequency and the vector-valued background field. For a purely timelike background, the refractive indices are real. Their associated propagation modes are circularly polarized and exhibit birefringence. For a purely spacelike background, one refractive index is always real and the other can be complex. The circularly polarized propagating modes may exhibit birefringence and dichroism (associated with absorption). Subsequently, we examine a dimension-five MCFJ-type electrodynamics, previously scrutinized in the literature, in a continuous medium.  Following the same procedure, we find the refractive indices from a sixth-order dispersion equation. For a purely timelike background, three distinct refractive indices are obtained, one of them being real and two being complex. They are associated with two circularly polarized propagating modes that exhibit birefringence or dichroism, depending on the frequency range. Scenarios of propagation and absorption analogous to those found in dispersive dielectrics are also observed for purely spacelike background configurations. We conclude by comparing the dimension-three and five results and by emphasizing the richer phenomenology of the propagating modes in the higher-derivative model. Our results are applicable in the realm of Weyl semimetals.

\pacs{41.20.Jb, 03.50.De, 03.50.-z, 11.30.Cp}
\keywords{Electromagnetic wave propagation; Maxwell equations; Lorentz invariance violation}

\end{abstract}

\maketitle

\section{\label{section1}INTRODUCTION}

The dynamics of electromagnetic fields in continuous media is governed by the Maxwell equations, supplemented by constitutive relations~\cite{Jackson,Zangwill} that describe the response of the medium to external, applied electromagnetic fields. \textit{In vacuo} these relations simply read ${\bf{D}}=\epsilon_{0}{\bf{E}}$ and ${\bf{H}}=\mu_{0}^{-1} {\bf{B}}$, where $\epsilon_{0}$ and $\mu_{0}$ are the electric vacuum permittivity and magnetic permeability constant, respectively. The first constitutive relation takes into account the electric polarization in a dielectric medium, while the latter includes magnetization effects. For an isotropic medium, the constitutive relations are ${\bf{D}}=\epsilon{\bf{E}}$ and ${\bf{H}}=\mu^{-1} {\bf{B}}$ with scalar material parameters $\epsilon,\mu$ replacing $\epsilon_0,\mu_0$ governing the vacuum properties. More involved constitutive relations appear in two main scenarios: (i) anisotropic media, where electric permittivity and magnetic permeability become tensors (cf.~uniaxial and biaxial crystals~\cite{Landau, Bain, Fowles, Hecht, Kurmanov, Yakov}, Weyl semimetals~\cite{Halterman, Zu}, and magnetized materials~\cite{ Krupka1, Krupka2}); (ii) novel effects in matter described by extended constitutive relations that introduce additional electric and magnetic responses, encoded as linear functions of the type $\mathbf{D}=\mathbf{D}(\mathbf{E},\mathbf{B})$ and $\mathbf{H}=\mathbf{H}(
\mathbf{E},\mathbf{B})$, in general. This happens, for instance, in bi-isotropic media~\cite{Aladadi, Sihvola, Sihvola2, Nieves}, chiral materials~\cite{Hillion}, topological insulators~\cite{Li1, Urrutia, Urrutia2, Lakhtakia, Winder, Li}, relativistic electron gases \cite{Carvalho}, axion electrodynamics~\cite{Sekine, Tobar2, BorgesAxion}, and Lorentz-violating electrodynamics~\cite{Tobar1, Bailey}, as well.

Generalizations of electrodynamics including higher-derivative terms have also been conceived in the literature. First studies of electrodynamics in the presence of higher derivatives are ascribed to Bopp \cite{Bopp} in 1940, and to Podolsky~\cite{Podolsky1,Podolsky2} in 1942. This model implements a second-order derivative term $\theta^{2}\partial_{\alpha}F^{\alpha\beta} \partial_{\sigma}F^{\sigma\beta}$ into the Maxwell Lagrangian \textit{in vacuo}. The modified Maxwell-Podolsky equations, sometimes called Bopp-Podolsky equations, yield a photon mass term, proportional to the inverse of the Podolsky parameter $\theta$. Furthermore, this extension exhibits two dispersion relations, the usual one from Maxwell theory and a second one ascribed to a massive mode.

The constraint structure of this theory was investigated in \cite{Galvao} and its quantization was performed in \cite{Barcelos}. Further aspects of the Maxwell-Podolsky model were examined, including the problems of self-force and self-interaction \cite{Gratus,Zayats}, Green functions and classical solutions \cite{Lazar}, multipole expansion for fields in the static regime~\cite{Bonin}, symmetrization/conservation of the energy-momentum tensor \cite{Fan}, its consistency based on the BRST approach \cite{Dai}, quantum field theoretic properties \cite{Bufalo,Zambrano} as well as other aspects \cite{Granado}.

A further example for a generalization of Maxwell electrodynamics is provided by Lee-Wick electrodynamics~\cite{Lee-Wick-1,Lee-Wick-2}, which introduces the modification $F^{\mu\nu} \square F_{\mu\nu}$~\cite{Turcati, Turcati2, Turcati3}. The higher-derivative Lee-Wick term can arise as a quantum correction in models with a nonminimal coupling between the gauge and fermionic fields~\cite{Borges}.

In the past years, higher-derivative contributions have also been examined in the context of Lorentz-violating theories. The possibility of Lorentz invariance violation (LV) was proposed in the context of physics at the Planck scale such as strings~\cite{Kostelecky:1988zi}. Presently, the Standard-Model Extension (SME)~\cite{Colladay}, where fixed background tensor fields are coupled to the dynamical Standard-Model fields, is usually employed as the main framework to parameterize it. Lorentz violation in the electromagnetic sector of the SME occurs by means of a \textit{CPT}-odd or a \textit{CPT}-even term~\cite{KM, Escobar, Santos, Belich}. The \textit{CPT}-odd part is represented by the Carroll-Field-Jackiw (CFJ) contribution~\cite{CFJ,CFJ2,CFJ3,CFJ4,CFJ5,CFJ6,CFJ7}, which has found applications in condensed-matter systems that violate parity (\textit{P}) and time reversal (\textit{T}) symmetry~\cite{Qiu} as well as those endowed with the chiral magnetic effect~\cite{Fukushima, Kharzeev, Kharzeev1,Kharzeev2} and the anomalous Hall effect~\cite{Zyuzin}.

Nonminimal extensions of the SME were proposed including higher-derivative terms with mass dimensions greater than four (in natural units)~\cite{Kostelecky, Mewes, Schreck}. In this context, the Myers-Pospelov model~\cite{Myers, Marat} was a pioneering proposal focusing on a dimension-five contribution. Recently, classical aspects of a modified, higher-derivative electrodynamics \textit{in vacuo} were discussed in~\cite{Leticia1,Leticia2}, including the derivation of the gauge propagator, the dispersion relations as well as an analysis of causality, unitary, and stability of the modes. Profound analyses were accomplished for the Maxwell-Podolsky electrodynamics modified by \textit{CPT}-even, dimension-six terms~\cite{Leticia2} and for a \textit{CPT}-odd, dimension-five electrodynamics~\cite{Leticia1}. Some results of Ref.~\cite{Leticia1} were revisited and discussed in Ref.~\cite{Passos2}.  Nonminimal higher-derivative models have also been used to study the interaction energy between electromagnetic sources~\cite{Borges-Ferrari} and the thermodynamic properties of electrodynamic systems~\cite{Filho-Maluf} as well as in the context of Hor\v ava-Lifshitz electrodynamics~\cite{Passos} and radiative corrections~\cite{Ferrari}.

The plethora of nonminimal LV theories on the one hand and the optical properties of new materials \cite{Aladadi,Shibata} on the other hand is a strong motivation for investigating higher-derivative effects on the propagation of electromagnetic waves in dielectric substrates, including aspects of optical activity and dichroism. In this sense, the present work is devoted to analyzing the behavior of a continuous medium governed by a MCFJ-type electrodynamics in the absence and presence of higher-derivative terms.

This paper is outlined as follows. In Sec.~\ref{section2}, we briefly review the covariant description of electrodynamics in macroscopic materials recapitulating the definition of birefringence (double refraction) and dichroism. We will be considering simple matter as opposed to designed materials with highly peculiar properties such as metamaterials \cite{Shelby, Valanju, Kshetrimayum}. In Sec.~\ref{section-comparison-to-MCFJ-1}, we present aspects of the MCFJ electrodynamics in a ponderable medium, showing that the timelike CFJ background yields birefringence, while the spacelike one provides birefringence and dichroism. In Sec.~\ref{section4}, we discuss the higher-derivative MCFJ electrodynamics in continuous matter based on more involved scenarios. Finally, we present our main findings in Sec.~\ref{final-remarks}. Throughout the paper, we employ natural units with $\hbar=c=1$ unless otherwise stated. Furthermore, our signature choice for the Minkowski metric $\eta_{\mu\nu}$ is $(+,-,-,-)$.

\section{\label{section2} Electrodynamics in simple matter}

In a continuous medium, the electromagnetic properties are described by the Maxwell equations \cite{Jackson, Zangwill} combined with the constitutive relations. For a general linear, homogeneous, and anisotropic medium, the constitutive relations can be written as
\begin{subequations}
\label{general-constitutive1}
\begin{align}
\mathbf{D}& =\epsilon\cdot \mathbf{E}+\gamma\cdot \mathbf{B}\,, \label{constitutive1} \\[1ex]
\mathbf{H}& =\tilde{\gamma}\cdot \mathbf{E}+\mu^{-1}\cdot \mathbf{B}\,,  \label{constitutive2}
\end{align}
\end{subequations}
where $\epsilon$ and $\mu$ represent the electric
permittivity and magnetic permeability tensors~\cite{Yakov, Aladadi, Sihvola, Sihvola2, Nieves}, respectively. The
tensor $\gamma$ measures the magnetic contribution to
the electric displacement field $\mathbf{D}$, while $\tilde{\gamma}$
represents the electric contribution to the magnetic field $\mathbf{H}$. Regarding the structure of constitutive relations~(\ref{constitutive1}) and (\ref{constitutive2}), interesting scenarios of electromagnetic behavior may occur, e.g., in anisotropic media~\cite{Aladadi, Bain, Fowles, Kurmanov, Yakov,Sihvola, Sihvola2, Nieves}, Weyl semimetals~\cite{Halterman, Zu, Grushin:2012mt}, magnetized ferrites~\cite{Krupka1, Krupka2}, and in chiral media and topological insulators~\cite{Urrutia, Urrutia2, Lakhtakia}. Besides \eqref{general-constitutive1}, general constitutive relations for the current density, ${\bf{J}}={\bf{J}}({\bf{E}}, {\bf{B}})$, can also be considered. As an example, a dielectric system endowed with a magnetic conductivity has recently been examined at the classical level~\cite{Pedro}, reporting interesting effects such as an induced electric conductivity, isotropic birefringence, and parity violation. A physical realization of the antisymmetric magnetic current examined in~\cite{Pedro} was addressed in Ref.~\cite{Kaushik}.

The constitutive relations in \eqref{general-constitutive1} can be naturally encoded in the Lagrange density formalism via
\begin{subequations}
\begin{equation}
\label{e1}
\mathcal{L}=-\frac{1}{4} G^{\mu\nu}F_{\mu\nu}-A_{\mu}J^{\mu}\,,
\end{equation}
with the four-potential $A_{\mu}$, the electromagnetic field strength tensor $F_{\mu\nu}=\partial_{\mu}A_{\nu}-\partial_{\nu}A_{\mu}$, and an external, conserved four-current $J^{\mu}$. Furthermore, the antisymmetric tensor $G^{\mu\nu}$ is defined as~\cite{Yakov}
\begin{equation}
\label{e2}
G^{\mu\nu}\equiv\frac{1}{2}\chi^{\mu\nu\alpha\beta}F_{\alpha\beta}\,,
\end{equation}
\end{subequations}
with $\chi^{\mu\nu\alpha\beta}$ being the constitutive tensor that parameterizes the medium's response to the applied electromagnetic fields~\cite{Post}. The constitutive tensor satisfies the following symmetry properties:
\begin{subequations}
\label{e3}
\begin{align}
\label{eq:symmetry-property-chi-1}
\chi^{\mu\nu\alpha\beta}&=-\chi^{\nu\mu\alpha\beta}\,, \\[1ex]
\label{eq:symmetry-property-chi-2}
\chi^{\mu\nu\alpha\beta}&=-\chi^{\mu\nu\beta\alpha}\,, \\[1ex]
\label{eq:symmetry-property-chi-3}
\chi^{\mu\nu\alpha\beta}&=\chi^{\alpha\beta\mu\nu}\,,
\end{align}
\end{subequations}
compatible with the symmetries of the field strength tensor. The Euler-Lagrange equation applied to the Lagrangian of \eqref{e1} (a complete derivation is presented in Appendix~\ref{AppendixA}) yields
\begin{equation}
\label{e4}
\partial_{\mu}G^{\mu\nu}=J^{\nu}\,.
\end{equation}
The homogenous Maxwell equations are obtained from the Bianchi identity valid for the curvature $F_{\mu\nu}$ of the principal \textit{U}(1) fiber bundle:
\begin{equation}
\label{e5}
\partial_{\mu}\tilde{F}^{\mu\nu}=0\,,\quad \tilde{F}^{\mu\nu}=\frac{1}{2}\epsilon^{\mu\nu\alpha\beta}F_{\alpha\beta}\,,
\end{equation}
which is why the latter are not affected by the presence of the medium.
A straightforward calculation from Eqs.~(\ref{e4}), (\ref{e5}) leads to the well-known Maxwell equations in simple matter, namely:
\begin{subequations}
\label{maxwell-equations}
\begin{align}
\nabla \cdot {\bf{D}}&=\rho\,, \label{maxwell-1} \displaybreak[0]\\[1ex]
 \nabla \times {\bf{H}} - \partial_{t}{\bf{D}}&={\bf{J}}\,,  \label{maxwell-2} \displaybreak[0]\\[1ex]
\nabla \cdot {\bf{B}}&=0\,, \label{maxwell-3} \displaybreak[0]\\[1ex]
 \nabla \times {\bf{E}}+\partial_{t}{\bf{B}}&=0\,,  \label{maxwell-4}
\end{align}
\end{subequations}
where the constitutive relations for the electric displacement field ${\bf{D}}$ and magnetic field ${\bf{H}}$ are given in \eqref{general-constitutive1}. The Maxwell equations  of~\eqref{maxwell-equations} and the constitutive relations given by \eqref{general-constitutive1} allow us to describe the dynamics of electromagnetic fields in simple matter.

In crystals and generic anisotropic media, the dispersion relations are given by the Fresnel equation~\cite{Zangwill}. The latter is obtained from algebraic manipulations of the Maxwell equations for continuous media whose properties are encoded in the permittivity tensor. In such a medium, the dielectric permittivity tensor is a function of the frequency $\omega$, the wave vector $\mathbf{k}$, and the (external) magnetic field $\mathbf{B}$ (or the magnetization, alternatively), $\epsilon_{ij}=\epsilon_{ij}(\omega, {\bf{k}}, {\bf{B}})$, and can be expanded as \cite{Shibata}
\begin{equation}
\label{eq:expansion}
\epsilon_{ij}(\omega, {\bf{k}}, {\bf{B}})=\epsilon^{0}_{ij}+\alpha_{ijl}k_{l}+\beta_{ijl}B_{l}+\gamma_{ijab}k_{a}B_{b}+\dots\,.
\end{equation}
The first term, $\epsilon^{0}_{ij}$, is the usual permittivity tensor of a dielectric. The second term, $\alpha_{ijl}k_{l}$, is a signature of spatial inversion symmetry breaking. It implies an optical activity that becomes manifest via linear birefringence or a rotation of the oscillation plane of linearly polarized light~\cite{Bain, Fowles}. The third term, $\beta_{ijl}B_{l}$, is associated with a fixed external magnetic field or magnetization and leads to a violation of time reversal symmetry. It gives rise to a magneto-optical activity via the Faraday or the Cotton-Mouton effect~\cite{Shibata}.

As already mentioned, a known consequence of the optical activity of a medium is linear birefringence, occurring when two circularly polarized modes of opposite chiralities, with refractive indices $n_{+}$ and $n_{-}$, respectively, have different phase velocities, $c/n_{+}$ and $c/n_{-}$. This property implies a rotation of the polarization plane of a linearly polarized wave. The latter phenomenon is quantified by the specific rotatory power $\delta$, which measures the rotation of the oscillation plane of linearly polarized light per unit traversed length in the medium. It is defined as
\begin{equation}
\label{eq:rotatory-power1}
\delta=-\frac{\omega}{2}[\mathrm{Re}(n_{+})-\mathrm{Re}(n_{-})]\,,
\end{equation}
where $n_{+}$ and $n_{-}$ are associated with left and right-handed circularly polarized waves, respectively.

Another interesting effect occurring in anisotropic crystals, dichroism, takes place when one polarization component is more strongly absorbed than the other. Obviously, this property is linked to the imaginary part of the refractive index. The difference of absorption of left and right-handed circularly polarized modes~\cite{Shibata} is given by the  dichroism coefficient [see Appendix~\ref{AppendixB} for the derivation of Eqs.~(\ref{eq:rotatory-power1}), (\ref{eq:dichroism-power1})]:
\begin{equation}
\label{eq:dichroism-power1}
{\delta}_{\mathrm{d}}=-\frac{\omega}{2}[\mathrm{Im}(n_{+})-\mathrm{Im}(n_{-})]\,.
\end{equation}
In the following, we examine two models of modified electrodynamics in continuous matter: the first one governed by the MCFJ Lagrangian and the second one by a higher-derivative MCFJ-type Lagrangian.

\section{\label{section-comparison-to-MCFJ-1}MCFJ model in a continuous medium}

In principle, MCFJ electrodynamics has connections with systems of chiral fermions, in particular, the chiral magnetic effect (CME), the anomalous Hall effect (AHE), and the anomalous generation of charge, besides birefringence effects.
In Ref.~\cite{Qiu}, the Maxwell equations \textit{in vacuo} modified by the CFJ background were obtained, which yields the terms ascribed to the CME and AHE. In this section, we examine aspects of the MCFJ electrodynamics embedded in a continuous medium. The latter plays a role for condensed-matter systems such as Weyl semimetals \cite{Yan:2016euz}. These novel materials are characterized by an even number of Weyl cones separated from each other in momentum space. In the vicinity of these cones, electrons behave as massless particles and have a certain Fermi velocity associated with them, whereupon their description via the Weyl equation is admissible. Having a microscopic realization of such a material at hand, it can be consistently described in the context of effective field theory via a $b$ term of the minimal SME fermion sector~\cite{Colladay}. The modified Dirac theory is frequently recast into the form~\cite{Grushin:2012mt,Behrends:2018qkj}
\begin{equation}
\label{eq:fermion-modified-b}
\mathcal{L}=\overline{\psi}\left[\gamma^{\mu}(\mathrm{i}\partial_{\mu}-b_{\mu}\gamma^5)-m\right]\psi\,,
\end{equation}
where $\psi$ is a Dirac spinor field, $\overline{\psi}$ its Dirac conjugate, $m$ the electron mass, $\gamma^{\mu}$ the standard Dirac matrices, and $\gamma^5$ is the chiral Dirac matrix. The vector-valued background field $b_{\mu}$ is known to catch the essential properties of a certain class of these materials. Integrating out the fermion fields implies an action for the electromagnetic fields. The latter decomposes into a \textit{CPT}-even part, which gives rise to a nontrivial permittivity and permeability of the system, as well as a \textit{CPT}-odd part corresponding to a CFJ term. We will come back to this point below.

The MCFJ Lagrange density in matter has the form
\begin{equation}
\mathcal{L}=-\frac{1}{4} G^{\mu\nu} F_{\mu\nu} - \frac{1}{4} \epsilon^{\beta\lambda\mu\nu} V_{\beta} A_{\lambda} F_{\mu\nu} - A_{\mu} J^{\mu}\,,\label{MCFJ-1}
\end{equation}
with the intrinsic vector-valued field $V^{\mu}$. Furthermore, $\epsilon^{\beta\lambda\mu\nu}$ is the Levi-Civita symbol in Minkowski spacetime fulfilling $\epsilon^{\beta\lambda\mu\nu}=-\epsilon_{\beta\lambda\mu\nu}$ and $\epsilon_{0123}=1$. The latter Lagrange density yields the following modified inhomogeneous Maxwell equations:
\begin{subequations}
\label{eq:mod-maxwell-MCFJ}
\begin{align}
\nabla \cdot {\bf{D}} - {\bf{V}} \cdot {\bf{B}} &= \rho\,,\label{mcfj-37} \\[1ex]
\nabla \times {\bf{H}} - \partial_{t} {\bf{D}} - V_{0} {\bf{B}} + {\bf{V}} \times {\bf{E}}&= {\bf{J}}\,, \label{mcfj-44}
\end{align}
\end{subequations}
where $\mathbf{V}$ is the spatial part of $V^{\mu}$ and the fields $\bf{D},\bf{H}$ fulfill the linear constitutive relations of \eqref{general-constitutive1}. Notice that the presence of the tensor $G^{\mu\nu}$ renders the Lagrangian~(\ref{MCFJ-1}) different from the one of Ref.~\cite{Qiu}, meaning that the modified Maxwell equations of~\eqref{eq:mod-maxwell-MCFJ} apply to a ponderable medium. Concerning the discrete symmetries, \textit{C} (charge conjugation), \textit{P}, and \textit{T}, it is worthwhile to recall that the CFJ term is \textit{CPT}-odd and the free part of the Lagrangian in~\eqref{MCFJ-1} can be written as:
\begin{align}
\mathrm{{\mathcal{L}}} &\supset \frac{1}{2}{\left( \mathbf{E}\cdot\mathbf{D}
-\mathbf{B}\cdot\mathbf{H}\right)}  \nonumber \\
&\phantom{{}\supset{}}+\frac{1}{2}\left[{{V^{0}\left(  \mathbf{A\cdot B}\right)  }}-{{A^{0}\left(  \mathbf{V\cdot B}\right)  }}
+{{\mathbf{V\cdot}\left(  \mathbf{A\times E}\right)  }}\right]\,. \label{mcfj-45}
\end{align}
In this sense, the pieces involving $V_{0}$ are \textit{P}-odd and \textit{T}-even, while the terms composed of $\bf{V}$ are \textit{P}-even and \textit{T}-odd, as properly shown in Tab.~\ref{table-behavior-CPT-MCFJ}. Thus, these terms can induce an optical activity of the medium (in the form of birefringence or dichroism).
\begin{table}[t]
	\caption{Behavior of the LV terms of the Lagrangian given by Eq.~(\ref{mcfj-45}) under \textit{C} (charge conjugation), \textit{P} (parity), and \textit{T} (time reversal).}
	\begin{centering}
		\begin{tabular}{ C{0.5cm}  C{0.7cm}  C{0.7cm} C{0.7cm} C{0.7cm}  C{1.4cm}  C{1.4cm} C{1.4cm}   C{0cm}}
			\toprule
			& $\textbf{E}$ & $\textbf{B}$ & $A_{0}$ & ${\bf{A}}$  & ${V}_{0} ({\bf{A}}\cdot {\bf{B}})$ & $A_{0} ({\bf{V}}\cdot {\bf{B}})$ &  $\mathbf{V\cdot}\left(  \mathbf{A\times E}\right)$ &  \\[2ex]
			\colrule
			\textit{C} & $-$          & $-$     & $-$  & $-$      & $+$      & $+$       & $+$       \\ [0.6ex]
			\textit{P} & $-$          & $+$     & $+$   & $-$        & $-$      & $+$       & $+$     \\[0.6ex]
			\textit{T} & $+$          & $-$     & $+$   & $-$        & $+$      & $-$    & $-$       \\ [0.6ex]
			\botrule
		\end{tabular}
	\end{centering}
	\label{table-behavior-CPT-MCFJ}
\end{table}

Another interesting aspect of MCFJ electrodynamics is that the term $V_{0} {\bf{B}}$ plays a significant role in a chiral magnetic current,
\begin{equation}
{\bf{J}}_{\mathrm{CME}}={\frac{e^{2}}{{4{\pi }^{2}}}}(\Delta\mu){\bf{B}}\equiv \Sigma {\bf{B}}\,,  \label{eq70}
\end{equation}
usually generated in chiral fermion systems~\cite{Pedro,Fukushima, Kharzeev, Kharzeev1,Kharzeev2}. Here,
$\Delta\mu \equiv {\mu}_{R}-\mu_{L}$ is also known as the chiral chemical potential and fermions of electric charge $\pm e$ are considered. In \eqref{eq70}, $\Sigma$ represents a chiral magnetic isotropic conductivity and plays a role equivalent to that of the timelike component $V_{0}$ in \eqref{mcfj-44}, as pointed out in Ref.~\cite{Qiu}. For the present analysis, we consider the usual constitutive relations. We set $\gamma_{ij}=\tilde{\gamma}_{ij}=0$ in \eqref{general-constitutive1}, implying
\begin{equation}
D^{i}=\epsilon _{ij}E^{j}\,,\qquad H^{i}=(\mu^{-1})_{ij}B^{j}\,, \label{DEB1}
\end{equation}
which can be restricted to the special scenario of an isotropic medium by choosing configurations proportional to the identity,
\begin{equation}
\label{isotropic-properties-1}
\epsilon_{ij}=\epsilon \delta_{ij}\,,\qquad (\mu^{-1})_{ij}=\mu^{-1} \delta_{ij}\,,
\end{equation}
where $\epsilon$ and $\mu$ are the electric permittivity and magnetic permeability constants, respectively. This approach is equivalent to taking the constitutive relations ${\bf{D}}=\epsilon {\bf{E}}$, and ${\bf{H}}=\mu^{-1}{\bf{B}}$.

In what follows, we implement the latter relations as well as ${\bf{J}}=\sigma {\bf{E}}$, where $\sigma$ is the Ohmic conductivity. Furthermore, we employ a plane-wave \textit{ansatz} for the fields, $\mathbf{E}=\mathbf{E}_{0}e^{\mathrm{i}(\mathbf{k}\cdot\mathbf{r}-\omega t)}$, $\mathbf{B}=\mathbf{B}_{0}e^{\mathrm{i}(\mathbf{k}\cdot\mathbf{r}-\omega t)}$, and similarly for ${\bf{D}}$ and ${\bf{H}}$ in \eqref{mcfj-44}. One then gets
\begin{equation}
\mathrm{i} {\bf{k}}\times {\bf{B}} + \mathrm{i} \mu \epsilon \omega {\bf{E}} - \mu V_{0} {\bf{B}} + \mu {\bf{V}} \times {\bf{E}} - \mu \sigma {\bf{E}}=0\,, \label{mod-ampere-2-2B}
\end{equation}
which can be simplified by using Faraday's law, yielding
\begin{subequations}
\begin{equation}
\left[\mathbf{k}^{2}\delta_{ij}-k_{i}k_{j}-\omega^{2} \mu \bar{\epsilon}_{ij} \right] E^{j}=0\,, \label{field-54}
\end{equation}
where we have defined
\begin{equation}
\bar{\epsilon}_{ij} \equiv \left(\epsilon + \mathrm{i} \frac{\sigma}{\omega}\right) \delta_{ij} - \frac{\mathrm{i}}{\omega^{2}} \epsilon_{iaj} \left(k_{a}V_{0} - \omega V_{a} \right)\,, \label{field-57}
\end{equation}
\end{subequations}
as an effective electric permittivity tensor. Here, $\epsilon_{ijk}$ is the Levi-Civita symbol in three dimensions. The second term on the right-hand side of \eqref{field-57}, $\epsilon_{iaj}k_{a}V_{0}$, is analogous to the tensor $\alpha_{ijl}k_{l}$ of \eqref{eq:expansion}, which breaks parity invariance. The third term, $\epsilon_{iaj}\omega V_{a}$, breaks time reversal invariance. Both are responsible for the optical activity of the medium, becoming manifest in birefringence, as we shall see.

For a continuous medium, we write $\mathbf{k}=\omega \mathbf{n}$, where $\mathbf{n}$ is a vector pointing along the direction of the wave vector and yielding the refractive index $n=+\sqrt{\mathbf{n}^2}$. To permit complex refractive indices, we take $+\sqrt{\mathbf{n}^2}$ instead of $|\mathbf{n}|$. The plus sign indicates, in principle, that we discard refractive indices with negative real parts, whenever such could occur. Composites with negative real parts of their refractive indices are called metamaterials~\cite{Shelby, Valanju, Kshetrimayum} and such possibilities will not be taken into account. Then, \eqref{field-54} becomes
\begin{subequations}
\begin{equation}
M_{ij}E^{j}=0\,, \label{field-56}
\end{equation}
with the tensor
\begin{equation}
M_{ij}= n^{2} \delta_{ij} - n_{i}n_{j} - \mu \bar{\epsilon}_{ij}\,, \label{field-56-1}
\end{equation}
where
\begin{equation}
\bar{\epsilon}_{ij}=\left(\epsilon + \mathrm{i} \frac{\sigma}{\omega}\right) \delta_{ij} - \frac{\mathrm{i}}{\omega} \epsilon_{iaj} \left(n_{a}V_{0} - V_{a} \right)\,.
\end{equation}
\end{subequations}
The nontrivial solutions for the electric field are obtained by requiring that the determinant of the matrix $M_{ij}$ vanish, $\mathrm{det}[M_{ij}]=0$, which yields the dispersion relations that describe wave propagation in the medium. The matrix $M_{ij}$ is explicitly given by
\begin{subequations}
\label{field-59}
\begin{align}
[M_{ij}]&=\mathcal{M}+\mathrm{i}\frac{\mu}{\omega}\mathcal{V}\,, \displaybreak[0]\\[2ex]
\label{eq:definition-curly-M}
\mathcal{M}&=\left[n^2-\mu\left(\epsilon+\mathrm{i}\frac{\sigma}{\omega}\right)\right]\mathds{1}_3-\begin{pmatrix}
n_1^2 & n_1n_2 & n_1n_3 \\
n_1n_2 & n_2^2 & n_2n_3 \\
n_1n_3 & n_2n_3 & n_3^2 \\
\end{pmatrix}\,, \displaybreak[0]\\[2ex]
\mathcal{V}&=\begin{pmatrix}
0 & V_3-V_0n_3 & V_0n_2-V_2 \\
V_0n_3-V_3 & 0 & V_1-V_0n_1 \\
V_2-V_0n_2 & V_0n_1-V_1 & 0 \\
\end{pmatrix}\,,
\end{align}
\end{subequations}
with the $(3\times 3)$ identity matrix $\mathds{1}_3$. The dispersion equation follows from $\mathrm{det}[M_{ij}] =0$:
\begin{subequations}
\label{field-69}
\begin{align}
0&=\tilde{\epsilon} \left(n^{2} - \mu\tilde{\epsilon}\right)^{2}-\frac{\mu}{\omega^{2}} \Big\{\mu\tilde{\epsilon} \left[V_{0}^{2} n^{2} + {\bf{V}}^{2} - 2 V_{0} ({\bf{n}}\cdot {\bf{V}} ) \right] \notag \\
&\phantom{{}={}}-{\bf{V}}^{2} n^{2} + ({\bf{n}}\cdot {\bf{V}})^{2}\Big\}\,,
\end{align}
where
\begin{equation}
\label{eq:definition-tilde-epsilon}
\tilde{\epsilon}=\epsilon+\mathrm{i}\frac{\sigma}{\omega}\,.
\end{equation}
\end{subequations}
We note that via the choices
\begin{equation}
\label{eq:effective-four-vectors}
\overline{p}^{\mu}\equiv\left(\sqrt{\tilde{\epsilon}}\,\omega,\frac{\mathbf{k}}{\sqrt{\mu}}\right)\,,\quad \overline{V}^{\mu}\equiv\left(\sqrt{\mu}\,V^0,\frac{\mathbf{V}}{\sqrt{\tilde{\epsilon}}}\right)\,,
\end{equation}
our \eqref{field-69} is equivalent to
\begin{equation}
\label{eq:MCFJ-medium-dispersion-equation}
\overline{p}^{4} + \overline{p}^{2} \overline{V}^{2} - (\overline{p} \cdot \overline{V} )^{2} =0\,.
\end{equation}
Alternatively, we can introduce an effective metric of the form
\begin{equation}
\label{eq:effective-metric}
\tilde{\eta}_{\mu\nu}\equiv\mathrm{diag}\left(\tilde{\epsilon},-\frac{1}{\mu},-\frac{1}{\mu},-\frac{1}{\mu}\right)\,,
\end{equation}
and write the dispersion equation as
\begin{align}
\label{eq:MCFJ-medium-dispersion-equation-effective}
0&=(p\cdot\tilde{\eta}\cdot p)^2 \notag \\
&\phantom{{}={}}+\frac{\mu}{\tilde{\epsilon}}\left[(p\cdot\tilde{\eta}\cdot p)(V\cdot\tilde{\eta}\cdot V)-(p\cdot\tilde{\eta}\cdot V)^2\right]\,.
\end{align}
\textit{In vacuo}, where $\epsilon=\mu=1$ and $\sigma=0$, the conventional four-momentum $p^{\mu}=(\omega, {\bf{k}})$ and the preferred spacetime direction  $V^{\mu}=(V^{0},\bf{V})$ satisfy
\begin{equation}
p^{4} + p^{2}V^{2} - (p \cdot V)^{2} =0\,,
\label{field-77}
\end{equation}
as expected. Equation~(\ref{field-77}) is the well-known dispersion equation of the MCFJ model \textit{in vacuo} \cite{CFJ,CFJ2}. Therefore, we interpret \eqref{eq:MCFJ-medium-dispersion-equation} as the dispersion equation for a generalized MCFJ theory in media. The four-vector $\overline{p}^{\mu}$ of \eqref{eq:effective-four-vectors} plays the role of an effective four-momentum that formally satisfies an analogous dispersion equation as \textit{in vacuo} when the preferred direction is replaced by $\overline{V}^{\mu}$. The possibility of expressing the dispersion equation in terms of the effective metric in \eqref{eq:effective-metric} and the conventional four-momentum $p^{\mu}$ is a different way of understanding this result. The presence of a medium described by the material parameters $\epsilon,\mu$, and $\sigma$ leads to electromagnetic waves obeying an analogous dispersion equation as in vacuum, but with the Minkowski metric replaced by an effective metric.

Now, let us have another look at MCFJ theory in the context of Weyl semimetals. As described in \cite{Grushin:2012mt}, integrating out the fermion fields of the effective field theory stated in \eqref{eq:fermion-modified-b} implies a modified electrodynamics described by the first two terms in \eqref{MCFJ-1}. Considering the realization of a Weyl semimetal studied in the latter reference leads to a particular choice of $V^{\mu}$ with $V^0=0$ and $\mathbf{V}$ pointing along the third spatial axis (compare the modified inhomogeneous Maxwell equations of their Eqs.~(31), (34) to our Eqs.~(\ref{mcfj-37}), (\ref{mcfj-44})). Although the CFJ term incorporates the most intriguing properties of such materials, we must also take into account that a Weyl semimetal (like any material) is characterized by a nontrivial permittivity (whereas the permeability is often simply to set 1). Thus, the optical response of such a material is very well described by a dispersion equation of the form of \eqref{eq:MCFJ-medium-dispersion-equation-effective} (cf.~Eq.~(36) in \cite{Grushin:2012mt}) being a formidable motivation for considering theories such as \eqref{MCFJ-1}. The author of the latter paper emphasizes that the presence of the CFJ term implies birefringence in Weyl semimetals.

In order to further understand some properties of MCFJ electrodynamics in a continuous dielectric medium with magnetic properties, we address two main scenarios: (i) a timelike and (ii) a spacelike background field $V^{\mu}$. We choose a non-Ohmic dielectric as a substrate, which implies $\sigma \mapsto 0$ in Eq.~(\ref{field-69}).

\subsection{\label{section-MCFJ-time-like-case}Purely timelike case}

\begin{figure}[b]
\begin{centering}
\includegraphics[scale=0.68]{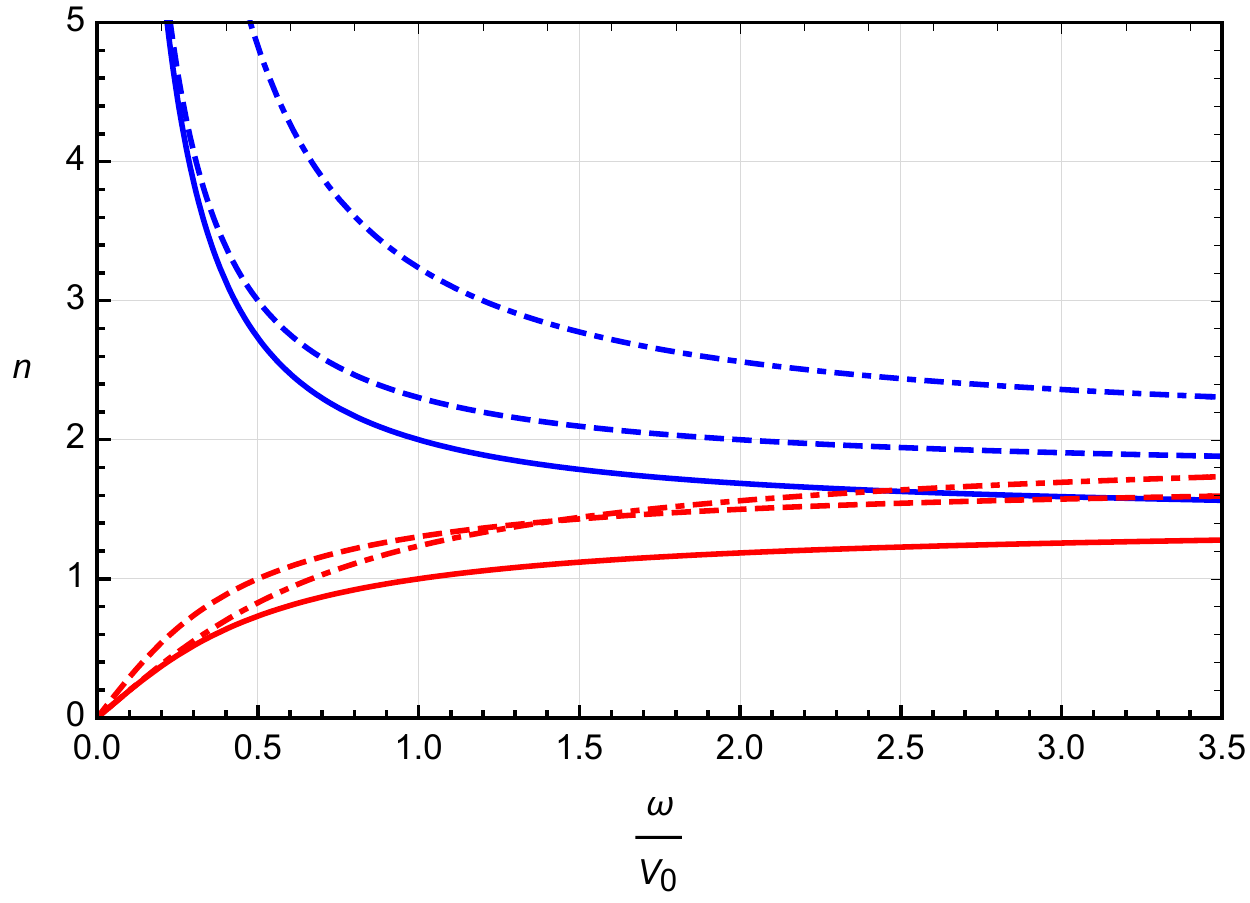}
\par\end{centering}
\caption{\label{plot-MCFJ-timelike}Refractive indices $n_{\pm}(\omega)$ of \eqref{field-82-1-1} in terms of $\omega/V_{0}$. Blue (monotonically decreasing) lines represent $n_{+}$, while red (monotonically increasing) lines depict $n_{-}$. For solid lines, $\mu=1$ and $\epsilon=2$; for dashed lines, $\mu=1$ and $\epsilon=3$; for dashed-dotted lines, $\mu=2$ and $\epsilon=2$.}
\end{figure}

For the purely timelike scenario, $V^{\mu}= (V^{0}, \bf{0})$, \eqref{field-69} reduces to
\begin{equation}
\left(n^{2} -\mu \epsilon \right)^{2} - \frac{\mu^{2} V_{0}^{2}}{\omega^{2}} n^{2} =0\,,
\label{field-79}
\end{equation}
which yields two distinct refractive indices:
\begin{subequations}
\begin{equation}
n_{\pm}^{2}=\mu\epsilon + \frac{\mu^{2} V_{0}^{2}}{2\omega^{2}} \pm \frac{\mu V_{0}}{\omega} \sqrt{ \mu \epsilon + \frac{\mu^{2} V_{0}^{2}}{4\omega^{2}}}\,,  \label{field-82}
\end{equation}
or equivalently
\begin{equation}
n_{\pm}=\sqrt{\mu \epsilon + \frac{\mu^{2} V_{0}^{2}}{4\omega^{2}}} \pm \frac{\mu V_{0}}{2\omega}\,. \label{field-82-1-1}
\end{equation}
\end{subequations}
The latter result is in accordance with the refractive index given in Eq.~(25) of Ref.~\cite{Pedro} for the diagonal isotropic magnetic conductivity tensor. This is an expected correspondence, since one knows that $V_{0}$ plays the role of a ``magnetic conductivity,'' as remarked below \eqref{eq70}. Note that $n_{\pm}$ are real and positive, allowing both modes to propagate for any frequency, so that an absorbing behavior is not observed here. Furthermore, in the limit of high frequencies, \eqref{field-82-1-1} provides $n_{\pm} \mapsto \sqrt{\mu\epsilon}$, recovering the refractive index of a medium with electric permittivity $\epsilon$ and magnetic permeability $\mu$, as described in the context of Maxwell electrodynamics. This behavior is illustrated in Fig.~\ref{plot-MCFJ-timelike}, which depicts the refractive indices (\ref{field-82-1-1}) in terms of the dimensionless parameter $\omega /V_{0}$ for some values of $\mu$ and $\epsilon$. The mode associated with $n_{+}$ exhibits anomalous dispersion, meaning that $\mathrm{d}n_{+}/\mathrm{d}\omega<0$, while $n_{-}$ is characterized by normal dispersion.
In order to examine the polarization state of the propagation modes, we first rewrite \eqref{field-79},
\begin{equation}
n^{2}-\mu\epsilon = \pm \frac{\mu V_{0}}{\omega} n\,.
\label{field-87}
\end{equation}
We employ the latter in \eqref{field-59}, whereupon the condition $M_{ij}E^{j}=0$ yields
\begin{equation}
{\bf{E}}_{\pm}=\frac{1}{\sqrt{2}n\sqrt{n^2-n_1^2}}\begin{pmatrix}
n^{2}-n_{1}^{2} \\
\mp\mathrm{i}n_{3}n-n_1n_2 \\
\pm\mathrm{i}n_{2}n-n_1n_3 \\
\end{pmatrix}\,.  \label{field-105}
\end{equation}
Considering the special choice
\begin{equation}
\label{field-146}
\bf{n}=\begin{pmatrix}
0 \\
0 \\
n_{3} \\
\end{pmatrix}\,,
\end{equation}
the normalized electric fields obtained from \eqref{field-105} are
\begin{align}
{\bf{E}}_{\pm} = \frac{1}{\sqrt{2}} \begin{pmatrix}
1 \\
\mp \mathrm{i} \\
0
\end{pmatrix}\,.
\label{field-117}
\end{align}
A polarization is defined to be right-handed (left-handed) if the polarization vector of a plane wave rotates along a circle in clockwise (counterclockwise) direction when the observer is facing into the incoming wave \cite{Jackson,Zangwill}. Therefore, $\mathbf{E}_{-}$ is interpreted as a left-handed and $\mathbf{E}_{+}$ as a right-handed circular polarization vector, respectively. These are associated with the distinct refractive indices $n_{-}$ and $n_{+}$ of \eqref{field-82-1-1} that imply different phase velocities of the physical modes giving rise to a rotation of the polarization plane of a linearly polarized wave. The implied birefringence is measured by the specific rotatory power [see the definition of \eqref{eq:rotatory-power1} and Appendix~\ref{AppendixB}], here written as
\begin{equation}
\label{eq:rotatory-power2}
\delta=-\frac{\mu V_{0}}{2}\,,
\end{equation}
which is a frequency-independent result dependent on the timelike component $V_{0}$ of the LV background. This nondispersive rotatory power differs from the rotatory power of a typical birefringent crystal, which increases with the frequency, as indicated by \eqref{eq:rotatory-power1} for constant refractive indices. As the refractive indices of \eqref{field-82-1-1} are real, there is no optical dichroism caused by $V_{0}$.

\subsection{\label{section-MCFJ-space-like-case}Purely spacelike case}

For the purely spacelike case, $V^{\mu}= (0,{\bf{V}})$, that is,  $V^{0}=0$ and ${\bf{V}}\neq 0 $, \eqref{field-69} yields
\begin{align}
\epsilon (n^{2}-\mu\epsilon)^{2} - \frac{\mu}{\omega^{2}} \left[ \mu \epsilon {{\bf{V}}}^{2} - n^{2} {\bf{V}}^{2} + ({\bf{n}\cdot \bf{V}})^2 \right] =0\,.
\label{field-129}
\end{align}
Implementing ${\bf{n}}\cdot {\bf{V}} = n | {\bf{V}}| \cos\theta$, one finds
\begin{subequations}
\begin{equation}
(n^{2}-\mu\epsilon)^{2} -\frac{\mu^{2}}{\omega^{2}} | {\bf{V}}|^{2} \alpha^{2}=0\,,
\label{field-134}
\end{equation}
where we defined
\begin{equation}
\alpha^{2}\equiv 1- \frac{n^{2}}{\mu\epsilon}\sin^{2}\theta\,.
\label{field-135}
\end{equation}
\end{subequations}
The two refractive indices (squared) read
\begin{align}
\label{eq:refractive-index}
n_{\pm}^2&=\mu\epsilon-\frac{\mu\mathbf{V}^2}{2\epsilon\omega^2}\sin^2\theta \notag \\
&\phantom{{}={}}\pm\frac{\mu|\mathbf{V}|}{2\epsilon\omega^2}\sqrt{4\epsilon^2\omega^2\cos^2\theta+\mathbf{V}^2\sin^4\theta}\,.
\end{align}
It is useful to analyze two special configurations: (i) the perpendicular case where ${\bf{n}}\cdot {\bf{V}} =0 $ and $\sin^{2}\theta=1$, (ii) the longitudinal case where ${\bf{n}} \cdot {\bf{V}} = \pm n | {\bf{V}}|$ and $\sin^{2}\theta=0$.

In order to examine the propagation modes, let us choose coordinates such that \eqref{field-146} holds, whereupon \eqref{field-59} simplifies as
\begin{equation}
[M_{ij}]= \begin{pmatrix}
n^{2}-\mu\epsilon  &&  \mathrm{i}\frac{\mu}{\omega} V_{3}  && - \mathrm{i} \frac{\mu}{\omega} V_{2} \\
\\
- \mathrm{i}\frac{\mu}{\omega} V_{3} && n^{2}-\mu\epsilon  &&  \mathrm{i} \frac{\mu}{\omega} V_{1} \\
\\
\mathrm{i}\frac{\mu}{\omega} V_{2} && - \mathrm{i} \frac{\mu}{\omega} V_{1} && -\mu\epsilon
\end{pmatrix}\,.
\label{field-147}
\end{equation}
Solving $M_{ij}E^{j}=0$, one obtains
\begin{subequations}
\label{eq:spacelike-configuration-generic-electric-field}
\begin{align}
{\bf{E}}_{\pm} = E_{0}\begin{pmatrix}
V_1V_2-\mathrm{i}\epsilon\omega V_3 \\
V_2^2-\epsilon^2\omega^2f(\alpha_{\pm}) \\
V_2V_3+\mathrm{i}V_1\epsilon\omega f(\alpha_{\pm}) \\
\end{pmatrix}\,,
\label{field-158}
\end{align}
where
\begin{equation}
f(\alpha)=1+\frac{\alpha^2-1}{\sin^2\theta}\,,\quad \alpha^{2}_{\pm} = 1-\frac{n^{2}_{\pm}}{\mu\epsilon}\sin^2\theta\,,
\label{field-135B}
\end{equation}
\end{subequations}
and $E_{0}$ is an appropriate normalization.

\subsubsection{\label{section-MCFJ-space-like-perpendicular}$\mathbf{V}$-perpendicular configuration}

\begin{figure}[t]
\begin{centering}
\includegraphics[scale=0.68]{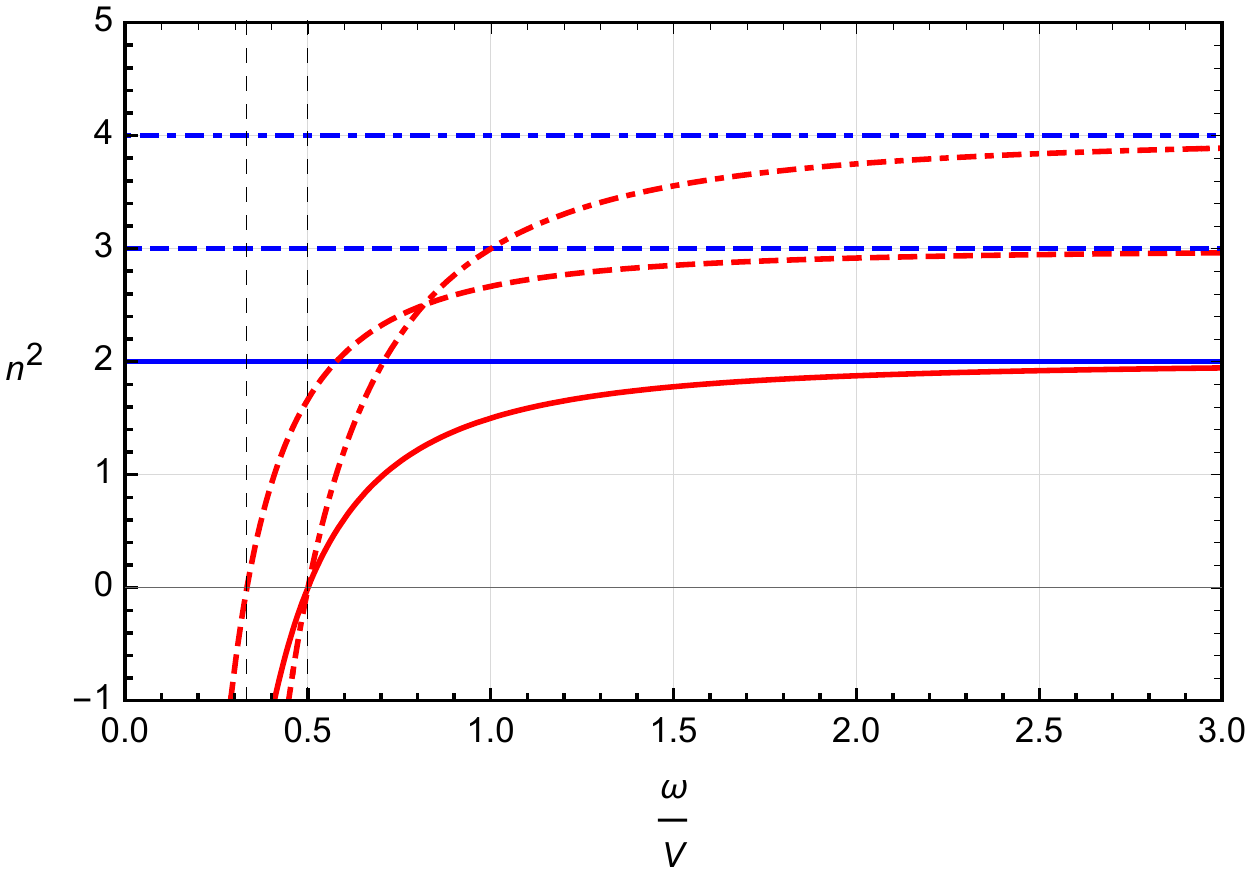}
\par\end{centering}
\caption{\label{plot-MCFJ-spacelike-perpendicular} Refractive indices $n^{2}_{\pm}(\omega)$ of \eqref{field-170} in terms of $\omega/V$ where $V=|{\bf{V}}|$. Blue (horizontal) lines represent $n^{2}_{+}$, while red (curved) lines depict $n^{2}_{-}$. For solid lines, $\mu=1$ and $\epsilon=2$; for dashed lines, $\mu=1$ and $\epsilon=3$; for dashed-dotted lines, $\mu=2$ and $\epsilon=2$.}
\end{figure}

Considering the perpendicular configuration with $\sin^{2}\theta=1$, the solutions of \eqref{field-129} for $n^{2}$ according to Eq.~(\ref{eq:refractive-index}) are
\begin{align}
n^{2}_{\pm}=\mu\epsilon+\frac{\mu {\bf{V}}^{2}}{2\epsilon\omega^{2}}(-1\pm 1)\,, \label{field-170}
\end{align}
that is
\begin{align}
n_{+}=\sqrt{\mu \epsilon}\,, \qquad n_{-}=\sqrt{\mu\epsilon-\frac{\mu\mathbf{V}^2}{\epsilon\omega^{2}}}\,.
\label{field-170B}
\end{align}
While $n_{+}$ is the standard refractive index of Maxwell electrodynamics in media, corresponding to $\alpha_{+}=0$, the refractive index $n_{-}$ is associated with $\alpha_{-}=|{\bf{V}}|/(\epsilon\omega)$, whereupon it is affected by the background. For $\omega<\omega_{-}$, we have $n_{-}^{2}<0$ and $n_{-}$ becomes purely imaginary, so that the corresponding mode no longer propagates. This defines the cutoff frequency,
\begin{align}
\omega_{-}= \frac{| {\bf{V}}|}{\epsilon}\,.  \label{field-170-1-1}
\end{align}
The general behavior of the refractive indices is depicted in Fig.~\ref{plot-MCFJ-spacelike-perpendicular}, where the squared refractive indices (\ref{field-170}) are plotted in terms of the dimensionless parameter $\omega/|{\bf{V}}|$. The horizontal dashed lines stand for $n^{2}_{+}$, which is constant for all frequencies. As for the mode associated with $n_{-}^{2}$, the vertical gray dashed lines (located at different $\omega_{-}/|{\bf{V}}|$ for each case) separate the absorption regime, $\omega < \omega_{-}$, from the propagation regime, $\omega > \omega_{-}$. Furthermore, in the limit of high frequencies, $n_{-}^{2} \mapsto n_{+}^{2}=\mu\epsilon$.

In order to examine the propagation modes, let us choose coordinates such that \eqref{field-146} holds. Then, a perpendicular background configuration is ${\bf{V}}=(V_{1}, V_{2},0)$. Due to $\alpha_+=0$ and $f(\alpha_+)=0$, \eqref{eq:spacelike-configuration-generic-electric-field} yields a linearly polarized, transverse mode,
\begin{equation}
\mathbf{E}_{+}=\frac{1}{|\mathbf{V}|}\begin{pmatrix}
V_{1} \\
V_{2} \\
0
\end{pmatrix}\equiv\hat{\mathbf{V}}\,,
\label{field-158-1-3}
\end{equation}
where $\hat{\mathbf{V}}$ is a unit vector pointing along the direction of $\mathbf{V}$. Also, inserting $\alpha_-$ into \eqref{eq:spacelike-configuration-generic-electric-field} provides another linearly polarized mode that has an additional longitudinal component:
\begin{align}
\bf{E}_{-}&= E_{0}^{(-)} \begin{pmatrix}
V_{2} \\
-V_{1} \\
\mathrm{i}(V_{1}^{2}+V_{2}^{2})/(\epsilon \omega) \\
\end{pmatrix} \notag \\
&=\tilde{E}_0^{(-)}\left(\hat{\mathbf{V}}\times\hat{\mathbf{n}}+\mathrm{i}\frac{|\mathbf{V}|}{\epsilon\omega}\hat{\mathbf{n}}\right)\,,  \label{field-158-1-4}
\end{align}
where $E_{0}^{(-)},\tilde{E}_0^{(-)}$ are properly chosen amplitudes and $\hat{\mathbf{n}}$ is the unit vector pointing along the propagation direction of \eqref{field-146}. Note that the longitudinal component is suppressed by the magnitude of the preferred direction $\mathbf{V}$ in comparison to the transverse part. For $V_{2}=0$ the behavior is even more transparent:
\begin{align}
{\bf{E}}_{+} = \begin{pmatrix}
1 \\
0 \\
0
\end{pmatrix}\,, \label{field-158-1-3B}
\qquad
{\bf{E}}_{-} = \tilde{E}_{0}^{(-)} \begin{pmatrix}
0 \\
-1 \\
\mathrm{i}  V_{1}/(\epsilon \omega)
\end{pmatrix}\,.
\end{align}
The structure of Eqs.~(\ref{field-158-1-3}), (\ref{field-158-1-4}) reveals immediately that $\mathbf{E}_+\cdot \mathbf{E}_-^{*}=\mathbf{E}_+^{*}\cdot \mathbf{E}_-=0$, i.e., both polarization vectors are orthogonal to each other.
The refractive indices (\ref{field-170B}) are associated with the linearly polarized modes of Eq.~(\ref{field-158-1-3B}). Although the vector $\mathbf{E}_-$ is composed of a transverse and a longitudinal component, as for polarization properties, it is interpreted as a linearly polarized mode and only its transverse component is taken into account.

If birefringence originates from two linearly polarized modes having different phase velocities, this property is not suitably characterized in terms of the usual rotatory power given by \eqref{eq:rotatory-power1}. Note that the latter is based on a decomposition of a linearly polarized mode into two circularly polarized ones of different chirality (see App.~\ref{AppendixB}). Instead, in the propagation regime, $\omega > \omega_{-}$, the phase shift developed between the propagating modes as a consequence of the distinct phase velocities is valuable to characterize birefringence (see Eq.~(8.32) in \cite{Hecht}):
\begin{equation}
\Delta=\frac{2\pi}{\lambda_0}d(n_{+}-n_{-})\,.
\label{phase-shift1}
\end{equation}%
Here, $\lambda_0$ is the wavelength of the electromagnetic radiation \textit{in vacuo} and $d$ corresponds to the thickness of the medium or the distance the wave travels in the medium. Starting from the refractive indices of Eq.~(\ref{field-170B}), the phase shift per unit length is
\begin{equation}
\frac{\Delta}{d}=\frac{2\pi}{\lambda_0}\sqrt{\mu \epsilon}\left(1-\sqrt{1-\frac{\mathbf{V}^{2}}{\epsilon^2\omega^{2}}}\right)\,,
\label{phase-shift2}
\end{equation}
which simplifies to
\begin{align}
\frac{\Delta}{d} &\simeq \frac{\pi\sqrt{\mu\epsilon}}{\lambda_0\epsilon^{2}\omega^2} {\bf{V}}^{2}\,, \label{birefringence-3}
\end{align}
in the limit $|{\bf{V}}|/\omega \ll 1$. Notice that $n_{-}$ is real for $\omega >\omega_-$ or $| {\bf{V}}|/\omega < \epsilon$, so that in the limit $| {\bf{V}}|/\omega \ll 1$, the expression (\ref{phase-shift2}) remains real, justifying the result~(\ref{birefringence-3}). These findings indicate that birefringence is governed by the norm squared of the LV background vector ${\bf{V}}$ and depends quadratically on the inverse of the frequency $\omega$, as well. That dependence is neither observed in the purely timelike case (see \eqref{eq:rotatory-power2} for comparison) nor in usual crystals (see \eqref{phase-shift1}).

For $\omega<\omega_-$, $n_{-}$ becomes complex while $n_{+}$ remains real. Thus, absorption (only) occurs for the mode labeled with a minus sign. In this case, the absorption coefficient~\cite{Zangwill}, $\gamma=2\omega \mathrm{Im}(n)$, reads
\begin{equation}
\gamma=2\sqrt{\mu\epsilon}\omega\sqrt{-1+\frac{|{\bf{V}}|^{2}}{\omega^{2}\epsilon^{2}}}\,.
\label{dichroism-2}
\end{equation}
So the mode associated with \eqref{field-158-1-4} is absorbed, whereas the remaining mode given by \eqref{field-158-1-3} propagates without attenuation. Therefore, after traveling a certain distance in such a medium, only the mode of~\eqref{field-158-1-3} will survive.

\subsubsection{\label{section-MCFJ-space-like-longitudinal}$\mathbf{V}$-longitudinal configuration}

\begin{figure}[t]
\begin{centering}
\includegraphics[scale=0.68]{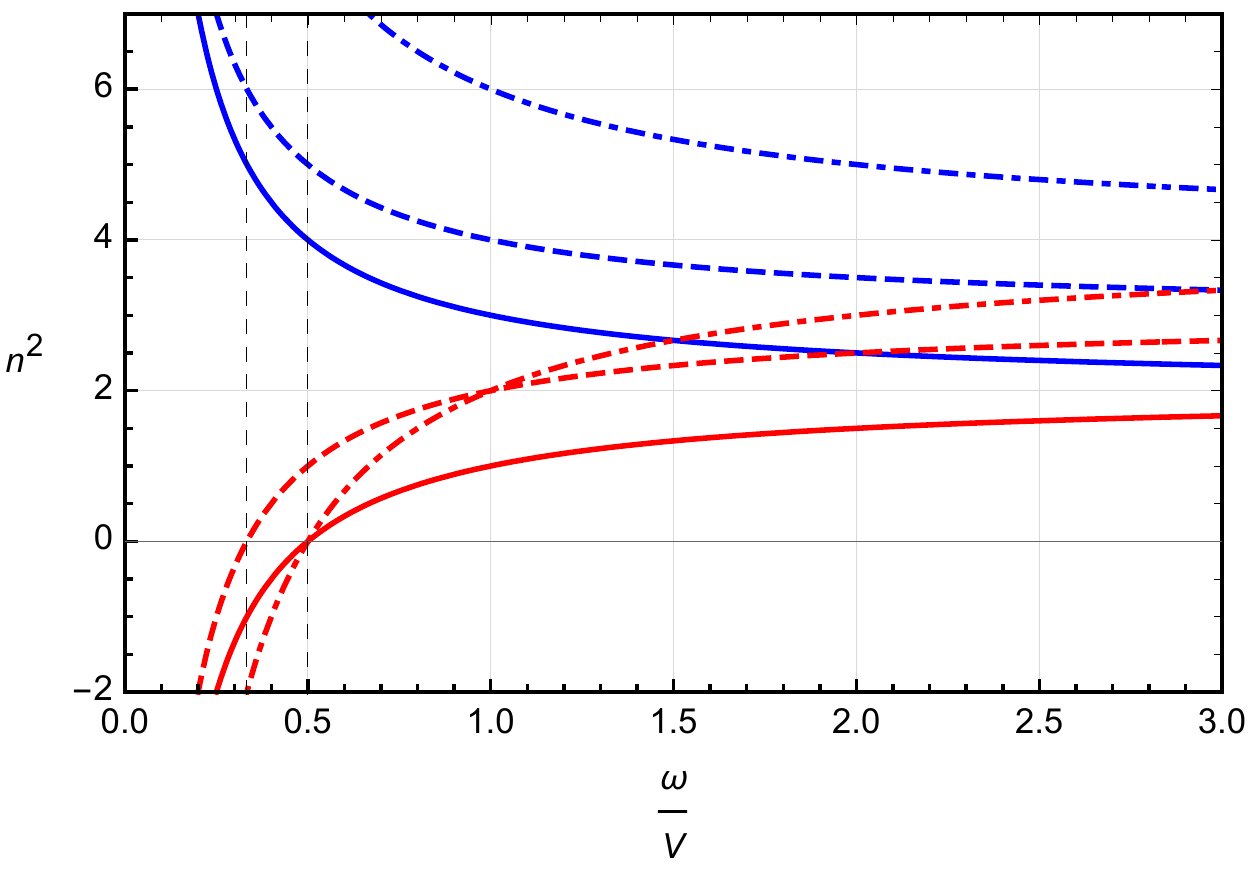}
\par\end{centering}
\caption{\label{plot-MCFJ-spacelike-longitudinal} Refractive indices $n^{2}_{\pm}(\omega)$ of \eqref{field-173} in terms of $\omega /V$ where $V=|{\bf{V}}|$. Blue (monotonically decreasing) lines represent $n^{2}_{+}$, while red (monotonically increasing) lines depict $n^{2}_{-}$. For solid lines, $\mu=1$ and $\epsilon=2$; for dashed lines, $\mu=1$ and $\epsilon=3$; for dotted-dashed lines, $\mu=2$ and $\epsilon=2$. The vertical dashed lines, from left to right, are given by $\omega_{-}/ {V} = 1/3$ and $\omega_{-}/ V = 1/2$, respectively, with $\omega_-$ from \eqref{field-170-1-1}.}
\end{figure}

We now consider configurations where ${\bf{n}}\cdot {\bf{V}} = \pm n | {\bf{V}}|$ implying $\sin^{2}\theta=0$ and $\alpha^{2} = 1$ in \eqref{field-134}.
This means that ${\bf{n}}$ and $ {\bf{V}}$ point along the same direction, i.e., for $\mathbf{n}$ given by \eqref{field-146} we choose ${\bf{V}}= (0,0, {V}_{3} )$. Hence, based on Eq.~(\ref{eq:refractive-index}), the solutions of \eqref{field-134} for $n^{2}$ in this case are
\begin{equation}
n^{2}_{\pm} = \mu \epsilon \pm \frac{\mu | {\bf{V}}|}{\omega}\,. \label{field-173}
\end{equation}
Note that $n_{+}^{2}>0$, meaning that the mode associated with the refractive index $n_{+}$ propagates within the full frequency domain. On the other hand, the mode associated with $n_{-}$ just propagates for $\omega >\omega_{-}$, for which $n_{-}^{2}>0$. Here,  $\omega_{-}$ is the cutoff frequency of \eqref{field-170-1-1}.
This description is verified in Fig.~\ref{plot-MCFJ-spacelike-longitudinal}, where the refractive indices (\ref{field-173}) are depicted as functions of the dimensionless parameter $\omega / |{\bf{V}}|$. The modes associated with $n^{2}_{+}$ and $n^{2}_{-}$ exhibit anomalous and normal dispersion, respectively, recovering the standard value $n_{\pm}^{2} \rightarrow \mu\epsilon$ in the regime of high frequencies.

For the special choice of \eqref{field-146}, the longitudinal background is of the form ${\bf{V}}= (0,0, {V}_{3})$, so that $M_{ij}E^{j}=0$ provides
\begin{align}
{\bf{E}}_{\pm} &= \frac{1}{\sqrt{2}}\begin{pmatrix}
1 \\
\pm \mathrm{i} \\
0
\end{pmatrix}\,.
\label{field-166}
\end{align}
Here, ${\bf{E}}_{+}$ and ${\bf{E}}_{-}$ represent polarization vectors for left-handed and right-handed circularly polarized modes, respectively.
The two refractive indices of \eqref{field-173} also imply birefringence providing the following rotatory power:
\begin{align}
\delta = - \frac{\sqrt{\mu\epsilon}}{2} \omega \left(\sqrt{1+ \frac{| {\bf{V}}|}{\omega \epsilon}} - \sqrt{1-\frac{| {\bf{V}}|}{\omega \epsilon}}\right)\,. \label{birefringence-4}
\end{align}
In the limit $| {\bf{V}}|/\omega \ll 1$, the quantity $n_{-}$ remains real, which also implies a real rotatory power,
\begin{align}
\delta \simeq -\frac{1}{2}\sqrt{\frac{\mu}{\epsilon}}| {\bf{V}}|\,, \label{birefringence-5}
\end{align}
representing frequency-independent birefringence, similarly to \eqref{eq:rotatory-power2}.

On the other hand, for $|{\bf{V}}|/\omega>\epsilon$, $n_{-}$ becomes purely imaginary, while $n_{+}$ remains real. In this frequency regime, both modes are absorbed to a different degree. The latter is characterized by the dichroism coefficient defined in \eqref{eq:dichroism-power1}, which yields (cf.~\eqref{dichroism-2}):
\begin{align}
\delta_{\mathrm{d}} &= \frac{\sqrt{\mu\epsilon}}{2} \omega \sqrt{-1 + \frac{ | {\bf{V}}|}{\omega \epsilon}}\,.
\label{dichroism-longitudinal-MCFJ}
\end{align}
With the latter finding at hand, we bring our study of the essential properties of MCFJ theory in continuous media to a close.

\section{\label{section4} Higher-derivative dimension-five electrodynamics in matter}

After analyzing the properties of MCFJ theory in a material (see \eqref{MCFJ-1}), the next logical step is to construct and investigate an extension involving additional four-derivatives. Such extensions are naturally contained in the nonminimal (nongravitational) SME \cite{Kostelecky, Mewes, Schreck}, which is a comprehensive framework for the parameterization of Lorentz and \textit{CPT} violation in effective field theory in Minkowski spacetime. For the past two decades it has been the foundation of various experiments testing the fundamental spacetime symmetries \cite{Kostelecky:2008ts}. No signal of Lorentz violation \textit{in vacuo} has been found, so far. However, Lorentz violation can be considered as an intrinsic property of material media, which is why the SME is more than suitable as a base for representing certain material properties within a field theory setting and to even propose novel materials with unusual characteristics.

The electromagnetic sector of the nonminimal SME gives rise to a modified electrodynamics and is given by
\begin{align}
\mathcal{L}&=-\frac{1}{4} F_{\mu\nu}F^{\mu\nu} + \frac{1}{2} \epsilon^{\kappa\lambda\mu\nu}A_{\lambda} (\hat{k}_{AF})_{\kappa}F_{\mu\nu} \notag \\
&\phantom{{}={}}-\frac{1}{4} F_{\kappa\lambda} (\hat{k}_{F})^{\kappa\lambda\mu\nu}F_{\mu\nu}\,. \label{higher-1}
\end{align}
The \textit{CPT}-odd and \textit{CPT}-even operators, $(\hat{k}_{AF})_{\kappa}$ and $(\hat{k}_{F})^{\kappa\lambda\mu\nu}$, respectively, are the analogs of $(k_{AF})_{\kappa}$ and $(k_{F})^{\kappa\lambda\mu\nu}$ of the minimal SME. However, they involve nonminimal coefficients contracted with additional four-derivatives in the form of the following infinite operator series:
\begin{subequations}
\begin{align}
(\hat{k}_{AF})_{\kappa} &= \sum_{d \text{ odd}} (k_{AF}^{(d)})_{\kappa}^{\phantom{1} \alpha_{1}\dots\alpha_{(d-3)}} \partial_{\alpha_{1}}\dots \partial_{\alpha_{(d-3)}} \,, \label{higher-2} \\[1ex]
(\hat{k}_{F})^{\kappa\lambda\mu\nu} &= \sum_{d \text{ even}} (k_{F}^{(d)})^{\kappa\lambda\mu\nu \alpha_{1}\dots\alpha_{(d-4)}}\partial_{\alpha_{1}}\dots \partial_{\alpha_{(d-4)}}\,, \label{higher-3}
\end{align}
\end{subequations}
where $d$ is the mass dimension of the tensor field operator that a certain coefficient is contracted with. Besides, $(4-d)$ is the mass dimension of the associated controlling coefficients $(k_{AF}^{(d)})_{\kappa}^{\phantom{1} \alpha_{1}\dots\alpha_{(d-3)}}$ and $(k_{F}^{(d)})^{\kappa\lambda\mu\nu \alpha_{1}\dots\alpha_{(d-4)}}$. The Lorentz indices $\alpha_{i}$ are contracted with additional spacetime derivatives.

We are interested in the \textit{CPT}-odd dimension-five ($d=5$) extension, which is represented by a CFJ-like term of the form
\begin{subequations}
\begin{equation}
\frac{1}{2} \epsilon^{\kappa\lambda\mu\nu}A_{\lambda} (\hat{k}_{AF})_{\kappa}F_{\mu\nu}\,,  \label{higher-4}
\end{equation}
with
\begin{align}
(\hat{k}_{AF})_{\kappa}=(k_{AF}^{(5)})_{\kappa}^{\phantom{\kappa}\alpha_{1}\alpha_{2}} \partial_{\alpha_{1}} \partial_{\alpha_{2}}\,. \label{higher-5}
\end{align}
\end{subequations}
For our investigation, we will use the parameterization
\begin{align}
(k_{AF}^{(5)})_{\kappa}^{\phantom{\kappa}\alpha_{1}\alpha_{2}}= {U}_{\kappa} \eta^{\alpha_{1} \alpha_{2}}\,, \label{higher-6}
\end{align}
with the Lorentz-violating four-vector, $U_{\kappa}$, and the Minkowski metric tensor, $\eta^{\mu\nu}$. Using \eqref{higher-6}, the higher-derivative term becomes
\begin{align}
\frac{1}{2} \epsilon^{\kappa\lambda\mu\nu} A_{\lambda} U_{\kappa} \square\,, \label{higher-7}
\end{align}
where we have introduced the d'Alembertian $\square=\eta^{\alpha_{1}\alpha_{2}}\partial_{\alpha_{1}}\partial_{\alpha_{2}}$. The resulting higher-derivative Lagrangian,
\begin{equation}
\mathcal{L}=-\frac{1}{4} F^{\mu\nu} F_{\mu\nu} +\frac{1}{2} \epsilon^{\beta\lambda\mu\nu} U_{\beta} A_{\lambda} \square F_{\mu\nu} -A_{\mu}J^{\mu}\,,
\label{lagrangian3A}
\end{equation}
involves LV parameterized by the background vector, $U^{\mu}=(U^{0}, {\bf{U}} )$. Some classical aspects of this model were examined in Refs. \cite{Leticia1, Marat}.

In order to study the effects of this higher-derivative term on electromagnetic propagation in continuous matter, we take as a starting point the Lagrangian (\ref{lagrangian3A}), but employ the field strength tensor $G^{\mu\nu}$ in its kinetic term, as it occurs in \eqref{e1}. Thus, the Lagrangian of this new model is
\begin{equation}
\mathcal{L}= -\frac{1}{4} G^{\mu\nu} F_{\mu\nu} +\frac{1}{2} \epsilon^{\beta\lambda\mu\nu} U_{\beta} A_{\lambda} \square F_{\mu\nu} -A_{\mu}J^{\mu}\,,
\label{lagrangian3}
\end{equation}
where the tensor $G^{\mu\nu}$ is written in terms of the constitutive tensor $\chi^{\mu\nu\alpha\beta}$, defined in Eqs.~(\ref{e2}), (\ref{e3}). The latter provides a generalization of the electrodynamics of \eqref{lagrangian3A} in matter. One may expect a connection between this theory and a generalization of the modified Dirac theory given by \eqref{eq:fermion-modified-b} where additional derivatives are included in the second contribution. However, it is beyond the scope of the current paper to demonstrate such a connection explicitly. Thus, by using \eqref{lagrangian3A} we can take into consideration an additional energy-momentum dependence that goes beyond that of the CFJ term in matter.

The Lagrangian of Eq.~(\ref{lagrangian3}) involves a third-order derivative of the four-potential, which requires an associated Euler-Lagrange equation endowed with derivatives for field derivatives that are of the same order. In principle, the derivative order can be decreased by rewriting Eq.~(\ref{lagrangian3}) in the form
\begin{equation}
\mathcal{L}=-\frac{1}{4} G^{\mu\nu} F_{\mu\nu} - \frac{1}{2} \epsilon^{\beta\lambda\mu\nu} U_{\beta} (\partial_{\eta} A_{\lambda}) \partial^{\eta} F_{\mu\nu} - A_{\mu}J^{\mu}\,.
\label{lagrangian4}
\end{equation}
As for the Lagrangian of Eq.~(\ref{lagrangian4}), it is enough to consider the Euler-Lagrange equation involving derivatives for second-order derivatives of the fields, that is,
\begin{equation}
\frac{\partial \mathcal{L}}{\partial A_{\kappa}}-\partial_{\rho} \left( \frac{\partial \mathcal{L}}{\partial (\partial_{\rho} A_{\kappa})} \right)+\partial_{\alpha}\partial_{\rho}  \left( \frac{\partial \mathcal{L}}{\partial (\partial_{\rho}\partial_{\alpha} A_{\kappa})} \right)=0\,. \label{second-order-derivative-lagrange-equations}
\end{equation}
Applying the latter to \eqref{lagrangian4} yields
\begin{equation}
\partial_{\rho}G^{\rho\kappa} + \epsilon^{\beta\kappa\mu\nu} U_{\beta} \square F_{\mu\nu} = J^{\kappa}\,.
\label{mod-maxwell-2}
\end{equation}
In this scenario, the modified Gauss's and Amp\`ere's laws are
\begin{subequations}
\begin{align}
\nabla \cdot {\bf{D}} + 2 \square ({\bf{U}} \cdot {\bf{B}} ) &= \rho\,, \label{mod-gauss-2} \displaybreak[0]\\[1ex]
\nabla \times {\bf{H}} - \partial_{t} {\bf{D}} + 2 \square {U}_{0} {\bf{B}} - 2 \square ( {\bf{U}}\times {\bf{E}} ) &= {\bf{J}}\,, \label{mod-ampere-2}
\end{align}
\end{subequations}
respectively. These modified inhomogeneous Maxwell equations can describe new effects on the propagation of electromagnetic waves in continuous media characterized by the constitutive tensor $\chi^{\mu\nu\alpha\beta}$. In the forthcoming sections we obtain the dispersion relations and study the behavior of refractive indices and propagating modes for a medium characterized by the usual constitutive relations, ${\bf{D}}=\epsilon{\bf{E}}$ and ${\bf{H}}=\mu^{-1} {\bf{B}}$.

With regards to the discrete symmetries, the background ${U}^{\mu}$ in the Lagrangian of Eq.~(\ref{lagrangian3}) behaves in the very exact way as $V^{\mu}$ does in Eq.~(\ref{MCFJ-1}), since the two \textit{CPT}-odd terms differ from each other by the presence of the second-order differential operator, $\square$, that is even under the discrete symmetries \textit{P} and \textit{T}. In fact, by simple inspection, one finds that the terms involving the timelike coefficient, $U_{0}$, are \textit{P}-odd, \textit{C}-even, \textit{T}-even, and \textit{PT}-odd, while the contributions with ${\bf{U}}$ are \textit{P}-even, \textit{C}-even, \textit{T}-odd, and \textit{PT}-odd, as shown in Tab.~\ref{table-behavior-CPT}. This means that the terms proportional to $U_{0}$ and ${\bf{U}}$ will act as a source for optical activity (as well as birefringence) of the medium under study.
\begin{table}[t]
	\caption{Behavior of the LV terms in the Lagrangian of Eq.~(\ref{lagrangian3A}) under charge conjugation, parity transformation, and time reversal.}
	\begin{centering}
	\begin{tabular}{ C{0.4cm}  C{0.45cm}  C{0.45cm} C{0.6cm} C{0.6cm}  C{1.6cm}  C{1.6cm}  C{1.66cm}  C{0cm}}
			\toprule
			& $\textbf{E}$ & $\textbf{B}$ & $A_{0}$ & ${\bf{A}}$  & ${U}_{0} ({\bf{A}}\cdot \square{\bf{B}})$ & $A_{0} ({\bf{U}}\cdot \square{\bf{B}})$ & ${\bf{U}} \cdot ({\bf{A}}\times \square{\bf{E}})$ &   \\[2ex]
			\colrule
			\textit{C} & $-$          & $-$     & $-$  & $-$      & $+$      & $+$    & $+$       \\ [0.6ex]
			\textit{P} & $-$          & $+$     & $+$   & $-$        & $-$      & $+$    & $+$      \\[0.6ex]
			\textit{T} & $+$          & $-$     & $+$   & $-$        & $+$      & $-$    & $-$        \\ [0.6ex]
			\botrule
		\end{tabular}
	\end{centering}
	\label{table-behavior-CPT}
\end{table}

\subsection{\label{dispersion-relations-1}Dispersion relations}

As is commonly known, the Maxwell equations constitute one starting point for achieving the dispersion relations in electrodynamics. Taking the time derivative of \eqref{mod-ampere-2} and employing \eqref{maxwell-4}, one obtains
\begin{equation}
\partial_{t} \nabla \times {\bf{H}} - \partial_{t}^{2} {\bf{D}} - 2 \square  U_{0} (\nabla \times {\bf{E}}) - 2 \partial_{t}\square ( {\bf{U}}\times {\bf{E}} ) = \partial_{t} {\bf{J}}\,.
\label{mod-ampere-2-1}
\end{equation}
Using now the constitutive relations given in Eqs.~(\ref{DEB1}) and (\ref{isotropic-properties-1}) as well as ${\bf{J}}=\sigma {\bf{E}}$ and the plane-wave \textit{ansatz} for the fields, \eqref{mod-ampere-2-1} yields
\begin{subequations}
\begin{align}
\left[\mathbf{k}^{2} \delta_{ij} - k_{i}k_{j} -\omega^{2} \mu \bar{\epsilon}_{ij}(\omega) \right] E^{j}=0\,,
\label{mod-ampere-2-4}
\end{align}
where we have defined the effective permittivity tensor (cf.~\eqref{field-57})
\begin{align}
\bar{\epsilon}_{ij} (\omega)&\equiv \left( \epsilon +\mathrm{i} \frac{\sigma}{\omega} \right)\delta_{ij}-\frac{2\mathrm{i}}{\omega^{2}}(k^{2}-\omega^{2}) \nonumber \\
&\phantom{{}={}}\times \epsilon_{iaj}\left(\omega U_{a} - k_{a} U_{0} \right)\,.  \label{mod-ampere-2-5}
\end{align}
\end{subequations}
The latter quantity is interpreted as an extended frequency-dependent electric permittivity, which contains contributions stemming from the higher-derivative term. On the right-hand side of \eqref{mod-ampere-2-5}, the contribution involving $\epsilon_{iaj}\omega U_{a}$ violates time reversal invariance, while the term $\epsilon_{iaj} k_{a}U_{0}$ breaks parity invariance. Using $\mathbf{k}=\omega \mathbf{n}$, \eqref{mod-ampere-2-4} can now be cast into the form:
\begin{subequations}
\begin{equation}
M_{ij}E^{j}=0\,,
\label{mod-ampere-2-7}
\end{equation}
with the tensor $M_{ij}$ given by
\begin{equation}
M_{ij}=n^{2} \delta_{ij} - n_{i}n_{j} - \mu \bar{\epsilon}_{ij} (\omega)\,,
\label{mod-ampere-2-8}
\end{equation}
while the effective permittivity tensor now reads
\begin{align}
\label{iso1}
\bar{\epsilon}_{ij} (\omega)&=\left(\epsilon+\mathrm{i}\frac{\sigma}{\omega}\right) \delta_{ij}-2 \mathrm{i} \omega (n^{2}-1) \nonumber \\
&\phantom{{}={}}\times \epsilon_{iaj}\left(U_{a}-n_{a} U_{0} \right)\,.
\end{align}
\end{subequations}
It is important to note that although the medium has an isotropic electric permittivity $\epsilon \delta_{ij}$, anisotropy effects are generated by the background $U_{\mu}$, present in the off-diagonal components of $\bar{\epsilon}_{ij}(\omega)$ in \eqref{iso1}.

The matrix $M_{ij}$ in \eqref{mod-ampere-2-8} has the explicit form
\begin{subequations}
\label{iso-matrix-1}
\begin{equation}
[M_{ij}]=\mathcal{M}+2\mathrm{i}\mu\omega (n^{2}-1)\mathcal{W}\,,
\end{equation}
with $\mathcal{M}$ given by \eqref{eq:definition-curly-M} and
\begin{equation}
\mathcal{W}=\begin{pmatrix}
0 & U_{0}n_{3}-U_{3} & U_{2}-U_{0}n_{2} \\[1ex]
U_{3}-U_{0}n_{3} & 0 & U_{0}n_{1}-U_{1} \\[1ex]
U_{0}n_{2}-U_{2} & U_{1}-U_{0}n_{1} & 0
\end{pmatrix}\,.
\end{equation}
\end{subequations}
Evaluating $\mathrm{det}[M_{ij}]=0$ implies the dispersion equation
\begin{align}
\label{iso-dispersion-relation-1}
0&=\tilde{\epsilon}(n^{2}-\mu \tilde{\epsilon})^{2}-4(n^{2}-1)^{2} \mu \omega^{2} \notag \\
&\phantom{{}={}}\times \left\{\mu \tilde{\epsilon} \left[ U_{0}^{2}n^{2}+{\bf{U}}^{2}-2 U_{0} ( {\bf{n}}\cdot {\bf{U}} )\right]\right. \notag \\
&\phantom{{}={}}\left.\hspace{0.6cm}{}-{\bf{U}}^{2} {\bf{n}}^{2} + ({\bf{n}}\cdot {\bf{U}} )^{2} \right\}\,,
\end{align}
with $\tilde{\epsilon}$ stated in \eqref{eq:definition-tilde-epsilon}. We point out that by employing the four-momentum of \eqref{eq:effective-four-vectors}
as well as
\begin{equation}
\overline{U}^{\mu}\equiv\left(\sqrt{\mu}\,{U}^0,\frac{{\mathbf{U}}}{\sqrt{\tilde{\epsilon}}}\right)\,,
\end{equation}
we can cast the dispersion equation into the form
\begin{equation}
\label{eq:MCFJ-dim5-medium-dispersion-equation}
\overline{p}^4+4p^4\left[\overline{U}^2\overline{p}^2-(\overline{U}\cdot\overline{p})^2\right]=0\,.
\end{equation}
By consulting the effective metric of \eqref{eq:effective-metric}, the latter can also be expressed in terms of the conventional four-momentum $p^{\mu}$ and the preferred direction ${U}^{\mu}$ as follows:
\begin{align}
\label{eq:MCFJ-dim5-medium-dispersion-equation-effective}
0&=(p\cdot\tilde{\eta}\cdot p)^2+4(p\cdot\eta\cdot p)^2\frac{\mu}{\tilde{\epsilon}} \notag \\
&\phantom{{}={}}\times\left[({U}\cdot\tilde{\eta}\cdot {U})(p\cdot\tilde{\eta}\cdot p)-({U}\cdot\tilde{\eta}\cdot p)^2\right]\,.
\end{align}
Note that in contrast to the dispersion equation of MCFJ theory stated in \eqref{eq:MCFJ-medium-dispersion-equation}, the recent \eqref{eq:MCFJ-dim5-medium-dispersion-equation} cannot be written in terms of the effective four-momentum $\overline{p}^{\mu}$ only, but $p^{\mu}$ is necessary, as well. The reason for $p^{\mu}$ playing a role are the two additional four-derivatives contracted with the dimension-5 coefficients in \eqref{higher-6}. Equation~(\ref{eq:MCFJ-dim5-medium-dispersion-equation-effective}) also allows us to say that the propagation of modified electromagnetic waves in media is governed by two metrics: the Minkowski metric $\eta_{\mu\nu}$ and the effective metric $\tilde{\eta}_{\mu\nu}$ of \eqref{eq:effective-metric}. Thus, the dimension-5 MCFJ-type theory defined by \eqref{lagrangian3} could be called bimetric in this sense. We conclude that the structure of the dimension-5 MCFJ-type theory in media is quite different from that of the generalized MCFJ model in \eqref{lagrangian3}.

\textit{In vacuo}, the constitutive parameters read $\epsilon=1$, $\mu=1$, and $\sigma=0$. In this case, the dispersion equation in \eqref{iso-dispersion-relation-1} reduces to
\begin{align}
0&=\left(n^{2}-1\right)^{2}\Big\{1 - 4\omega^{2} \left[ {U}_{0}^{2} n^{2} - {\bf{U}}^{2}n^{2} +  {\bf{U}}^{2} \right. \nonumber \\
&\phantom{{}={}}\left.\hspace{1.6cm}{}+ ( {\bf{n}}\cdot {\bf{U}} )^{2} - 2 {U}_{0} ({\bf{n}}\cdot {\bf{U}} ) \right]\Big\}\,,
\label{iso3}
\end{align}
being conveniently simplified as
\begin{equation}
p^{4} \left\{1+4 \left[p^{2} {U}^{2} - ({U} \cdot p )^{2} \right] \right\}=0\,,
\label{iso4}
\end{equation}
with the four-momentum $p^{\mu}$ and the preferred direction ${U}^{\mu} = ({U}^{0}, {\bf{U}} )$. Notice that \eqref{iso4} recovers the dispersion equation obtained in Eq.~(23) of Ref.~\cite{Leticia1}, where this higher-derivative electrodynamics was examined \textit{in vacuo}. It is important to point out that the remarkable difference between \eqref{eq:MCFJ-dim5-medium-dispersion-equation} and \eqref{iso4} is ascribed to the presence of the continuous medium, since the dimension-five higher-derivative terms in the Lagrangians of Eqs.~(\ref{lagrangian3A}), (\ref{lagrangian3}) correspond to each other.

In what follows, we analyze the dispersion equation~(\ref{iso-dispersion-relation-1}) for the timelike and spacelike configurations of the vectorial background, ${U}^{\mu}$.

\subsection{\label{section-time-like-case}Purely timelike case}

Considering the purely timelike scenario for the background vector, ${U}_{0} \neq 0$ and ${\bf{U}}= {\bf{0}}$, and also $\tilde{\epsilon} \mapsto \epsilon$, which means that the medium does not have Ohmic conductivity (whereupon $\sigma=0$), \eqref{iso-dispersion-relation-1} is reduced to the form
\begin{subequations}
\begin{equation}
{\epsilon} (n^{2} - \mu {\epsilon})^{2} - 4 \mu^{2} \omega^{2} {U}_{0}^{2} {\epsilon} n^{2} (n^{2}-1)^{2}=0\,, \label{timelike1}
\end{equation}
implying
\begin{equation}
n^{2}-\mu {\epsilon}= \pm 2 \mu \omega {U}_{0} n (n^{2}-1)\,, \label{time3}
\end{equation}
or equivalently
\begin{equation}
\pm 2 \mu \omega {U}_{0} n^{3} - n^{2} \mp 2 \mu \omega {U}_{0} n + \mu \epsilon=0\,.  \label{time4}
\end{equation}
\end{subequations}
The latter equation is cubic in $n$ and has 3 (complex) solutions, in general, given as functions $n=n(\omega)$. These solutions extend to frequency domains defined in accordance with the sign of the discriminant of the cubic equation, written as
\begin{subequations}
\label{eq:discriminant}
\begin{equation}
\Delta= \frac{S}{2^{4} 3^{3} \mu^{3} \omega^{4} {U}_{0}^{4}}\,, \label{time12}
\end{equation}
with
\begin{equation}
S = -\epsilon - \mu \omega^{2} {U}_{0}^{2} \left[1+ 9 \mu \epsilon \left(2-3\mu\epsilon \right) + 16 \mu^{2} \omega^{2} {U}_{0}^{2} \right]\,. \label{time13}
\end{equation}
\end{subequations}
For a cubic polynomial equation, the sign of $\Delta$ helps us to identify the nature (real or complex) of the 3 solutions, in accordance with Tab.~\ref{tab:discriminant}.
\begin{table}[t]
	\caption{Sign of discriminant $\Delta$ of \eqref{eq:discriminant} and the nature of the roots (solutions) of \eqref{time4}.}
	\begin{centering}
		\begin{tabular}{ C{1.2cm}  C{7cm}  C{.01cm}  }
			\toprule
			Sign & Solutions &  \\[0.6ex]
			\colrule
			$\Delta > 0 $ & one real root and two complex conjugate roots \\[0.6ex]
			$\Delta \leq 0$ &  three real roots (with two or all three equal to each other if $\Delta = 0$) \\[0.6ex]
			\botrule
		\end{tabular}
	\end{centering}
	\label{tab:discriminant}
\end{table}
\begin{figure}[b]
\begin{centering}
\includegraphics[scale=0.68]{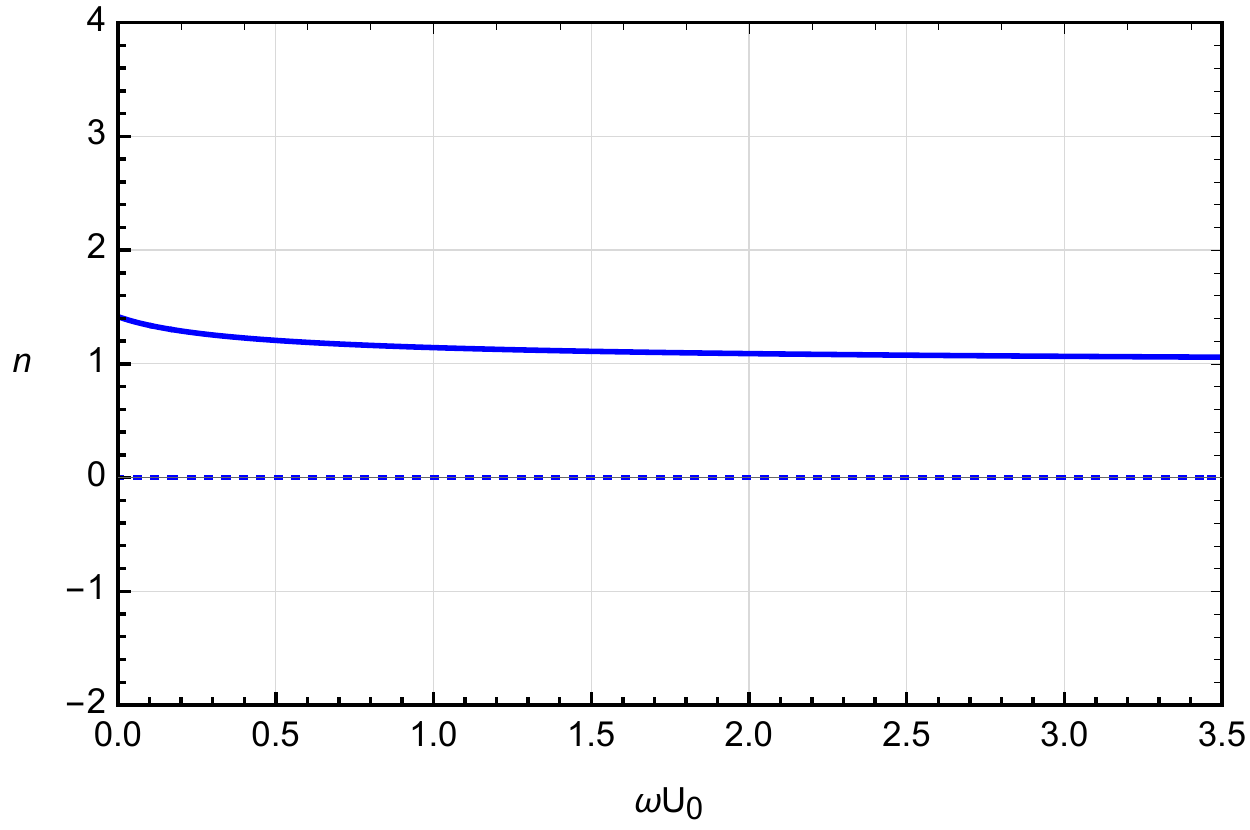}
\par\end{centering}
\caption{\label{figure-time-like-2} Plot of one real root of \eqref{time4}, the refractive index $n_{1}(\omega)$, in terms of $\omega {U}_{0}$. It is obtained by choosing the lower signs of~\eqref{time4}. The solid (dotted) line represents $\mathrm{Re}[n_{1}(\omega)]$ ($\mathrm{Im}[n_{1}(\omega)]$) where the latter vanishes.}
\end{figure}

Since the denominator of \eqref{time12} is positive, we only need to analyze the sign of the numerator, $S$. As $S$ is a function quartic in $\omega$, it is possible to find two roots that provide three frequency ranges for positive or negative values of $\Delta$. In this way, the relation $S=0$ establishes the critical values of frequencies (roots) that separate the absorption domain $S>0$ from the propagation domain $S<0$. Solving $S=0$, one achieves two roots for $\omega^2$ given by
\begin{align}
\omega^{2}_{\pm} &= \frac{1}{32\mu^{2} {U}_{0}^{2}} \bigg\{9\mu\epsilon \left(3\mu\epsilon-2\right)-1 \notag \\
&\phantom{{}={}}\hspace{1.5cm}\pm \sqrt{\mu\epsilon-1}(9\mu\epsilon-1)^{3/2}\bigg\}\,. \label{time14}
\end{align}
Thus, the three frequency ranges associated with two distinct scenarios are as follows:

\begin{itemize}

\item[i)] For $\omega_{-} < \omega < \omega_{+}$, one has $S>0$ and $\Delta>0$, so that \eqref{time4} yields one real function $n(\omega)$ and two complex functions $n(\omega)$.
\item[ii)] For $\omega<\omega_{-}$ or $\omega > \omega_{+}$, one has $S<0$ and $\Delta<0$, so that there are three real refractive indices $n(\omega)$.

\end{itemize}

The first domain describes absorption effects, whereas electromagnetic waves can freely propagate without attenuation in the second domain.
The sign of $S$ determines the real or complex nature of $n(\omega)$ in the corresponding frequency range. For a complex refractive index, we can write $n(\omega)=n'(\omega)+\mathrm{i}n''(\omega)$, where $\mathrm{Re}[n(\omega)]=n'(\omega)$ is the refractive index of the medium, and $\mathrm{Im}[n(\omega)]=n''(\omega)$ is associated with the medium's absorption coefficient $\alpha=2\omega n''(\omega)$~\cite{Zangwill}.

\begin{figure}[t]
\begin{centering}
\includegraphics[scale=0.68]{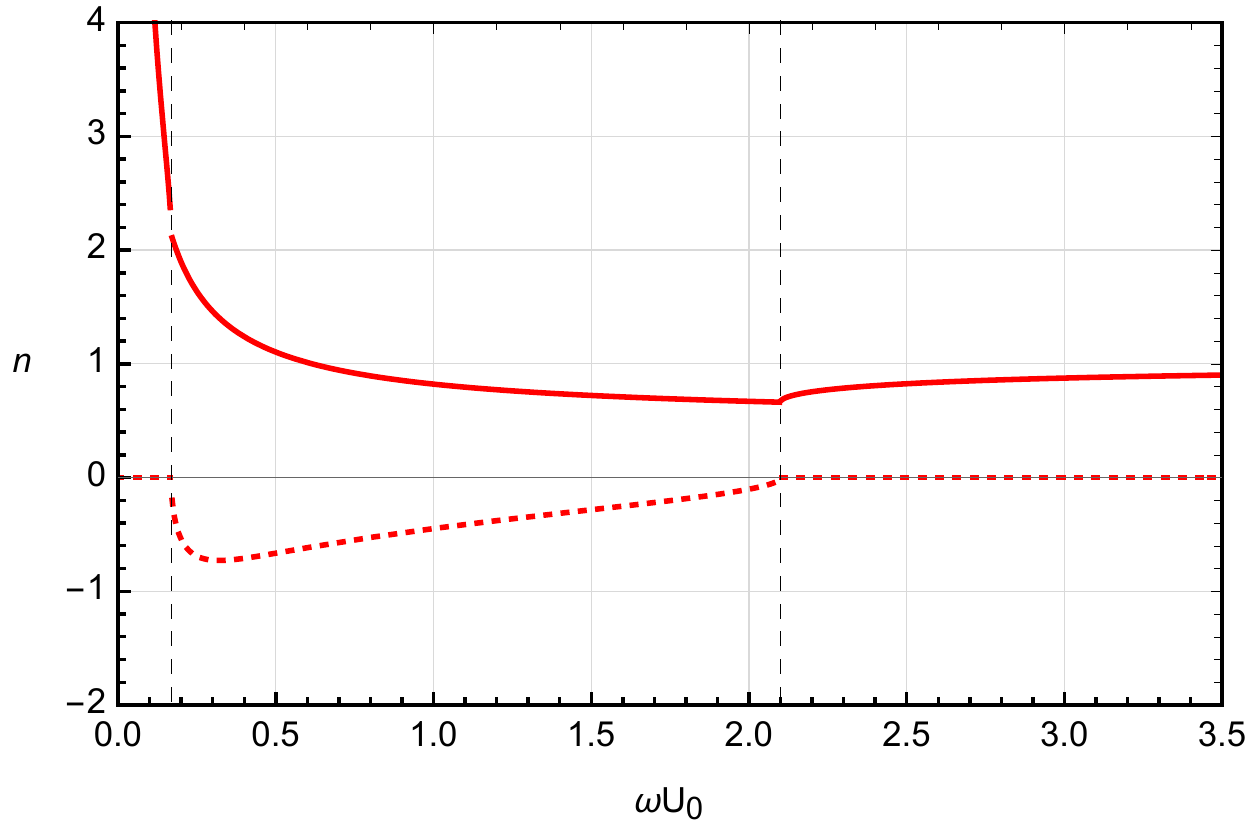}
\par\end{centering}
\caption{\label{figure-time-like-1} Plot of one complex root of \eqref{time4}, the refractive index $n_{2}(\omega)$, in terms of $\omega {U}_{0}$. It follows from \eqref{time4} with the upper signs taken into account. The solid (dotted) line represents $\mathrm{Re}[n_{2}(\omega)]$ ($\mathrm{Im}[n_{2}(\omega)]$).}
\end{figure}

Joining the above domains, we can conclude that:

\begin{itemize}

\item[a)] For $\omega<\omega_{-}$ there are three real solutions.
\item[b)] For $\omega_{-} < \omega < \omega_{+}$ two solutions become complex and the remaining one stays real.
\item[c)] For $\omega > \omega_{+}$ the three solutions become real again.

\end{itemize}

In general, propagation without attenuation is associated with real (positive) refractive indices, whereas absorption effects (damping of the amplitude of electromagnetic waves) are related to complex refractive indices. The modified electrodynamics defined by \eqref{lagrangian4} ascribes a conducting behavior to a dielectric substrate (with additional magnetic properties). For the particular scenario studied previously, electromagnetic waves propagate without being damped in the frequency range where the three solutions are real. In the range where complex solutions for $n(\omega)$ occur, both propagation and absorption (attenuation) is observed. These novel effects stem from the higher-derivative coupling of the background coefficient ${U}_{0}$ with the electromagnetic fields.

\begin{figure}[t]
\begin{centering}
\includegraphics[scale=0.68]{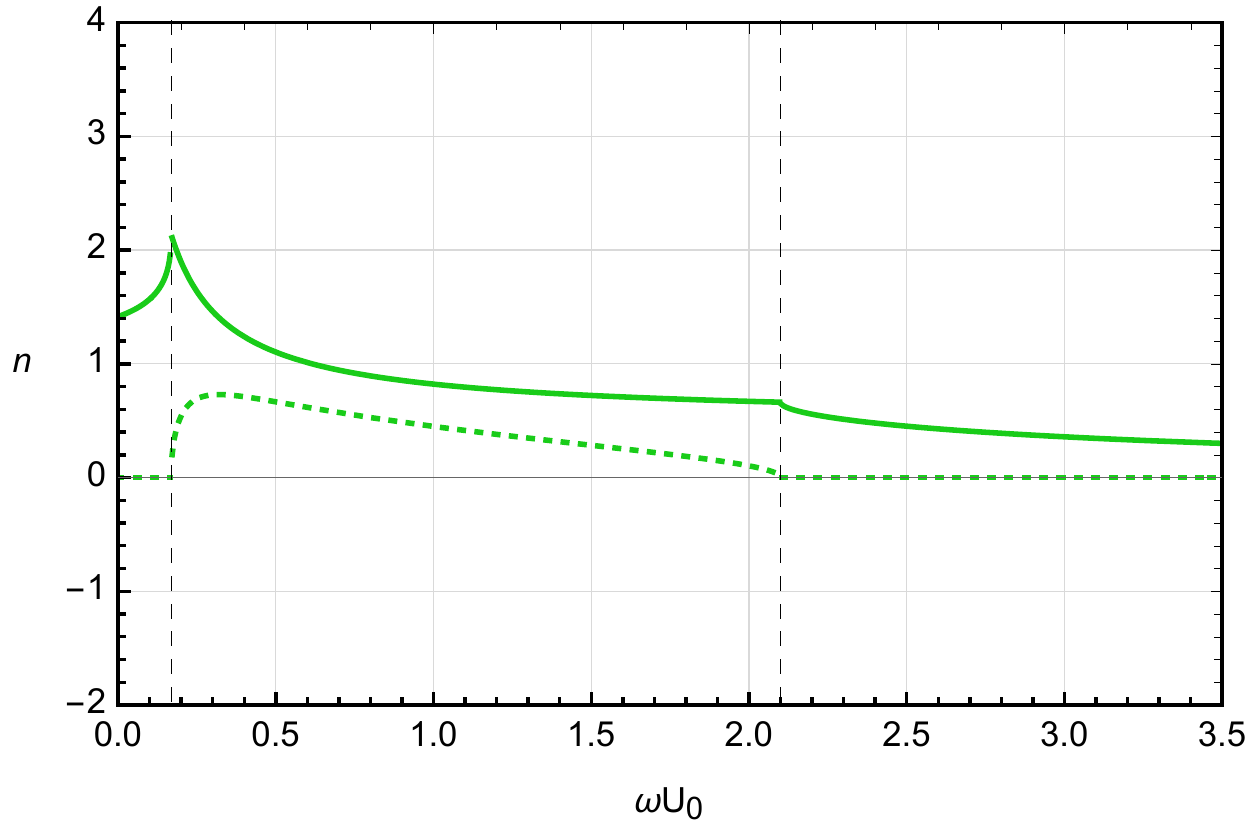}
\par\end{centering}
\caption{\label{figure-time-like-3} Plot of one complex root of \eqref{time4}, the refractive index $n_{3}(\omega)$, in terms of $\omega {U}_{0}$. It results from choosing the upper signs of \eqref{time4}. The solid (dotted) line depicts $\mathrm{Re}[n_{3}(\omega)]$ ($\mathrm{Im}[n_{3}(\omega)]$).}
\end{figure}

The refractive indices for a continuous medium with signal propagation described by \eqref{time4} are given by very intricate expressions (the roots of \eqref{time4}), which will not be stated here explicitly. We depict these three functions, $n_{i}(\omega)$, for $i=1,2,3$ in terms of the dimensionless parameter $\omega {U}_{0}$ for the special values $\mu=1$ and $\epsilon=2$. These plots are presented in Figs.~\ref{figure-time-like-2}, \ref{figure-time-like-1}, and \ref{figure-time-like-3}, where the solid (dotted) lines stand for the real (imaginary) part of $n(\omega)$. The refractive indices shown in the previous figures are characterized by positive real parts. The remaining three refractive indices, which follow from the generic sixth-order polynomial of Eq.~(\ref{timelike1}), have negative real parts.

\begin{figure}[h]
\begin{centering}
\includegraphics[scale=0.68]{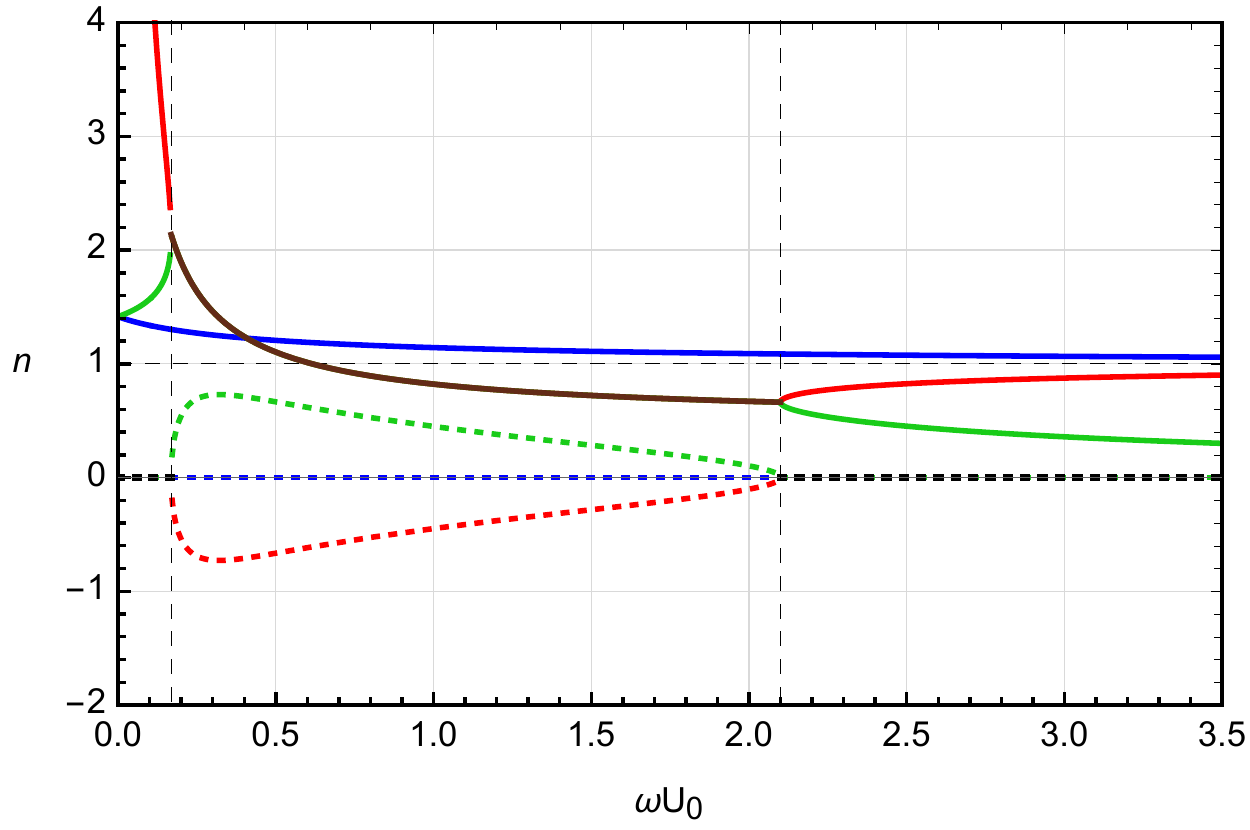}
\par\end{centering}
\caption{\label{figure-time-like-all-together} Compilation of complex refractive indices $n(\omega)$ from Figs.~\ref{figure-time-like-2}, \ref{figure-time-like-1}, and \ref{figure-time-like-3}. The solid lines illustrate the real parts of $n_{1}(\omega)$ (blue), $n_{2}(\omega)$ (red), and $n_{3}(\omega)$ (green). Their corresponding imaginary pieces are represented by dotted lines with the same colors. The solid brown line indicates that $\mathrm{Re}[n_{2}(\omega)]$ and $\mathrm{Re}[n_{3}(\omega)]$ lie on top of each other. Black dotted lines are employed whenever all three imaginary parts merge.}
\end{figure}

We notice that $n_{1}(\omega)$ is always real for all frequency ranges. The functions $n_{2}(\omega)$ and $n_{3}(\omega)$ become complex in the range $\omega_{-} < \omega < \omega_{+}$, in agreement with the previous analysis.
Combining all three plots in Fig.~\ref{figure-time-like-all-together}, we realize the full scenario described in items (a) -- (c) previously stated. The vertical dashed lines indicate the critical frequency values of Eq.~(\ref{time14}), namely $\omega_{-} {U}_{0}$ and $\omega_{+} {U}_{0}$, which define the transition between the ranges given in (a) -- (c). Another characteristic of Figs.~\ref{figure-time-like-1} and \ref{figure-time-like-3} are the discontinuities in the real parts of $n_{2}(\omega)$ and $n_{3}(\omega)$, at the frequencies $\omega_{\pm}$. Note that $n_{2}(\omega)$ and $n_{3}(\omega)$ become purely imaginary at these values.

As a final comment, we point out that the physical behavior described above only occurs for the higher-derivative electrodynamics of \eqref{lagrangian4} in matter. In fact, \textit{in vacuo}, \eqref{timelike1} would provide
\begin{equation}
(n^{2}-1)^{2}(1- 4 \omega^{2} {U}_{0}^{2} n^{2})=0\,, \label{timelike1-1}
\end{equation}
whose solutions are real, namely:
\begin{equation}
n=1, \quad n=\frac{1}{2\omega|{U}_{0}|}\,, \label{timelike1-3}
\end{equation}
meaning the absence of absorption effects \textit{in vacuo} (for this dimension-5 theory). This behavior can also be inferred directly from \eqref{time14}, since $\Delta \omega=\omega_{+}-\omega_{-} =0$, for $\mu=\epsilon=1$, corresponding to the disappearance of frequency ranges where absorption occurs.

Furthermore, the second refractive index of \eqref{timelike1-3} does not have a well-defined limit for $U_0\mapsto 0$. \textit{In vacuo}, such modes are sometimes called spurious and their occurrence is characteristic for higher-derivative theories (see, e.g., \cite{Schreck,Leticia2,Leticia1} for detailed investigations in the nonminimal electromagnetic sector of the SME). They can be interpreted as high-energy effects decoupling from the theory at low energies. However, a finite $U_0$ in macroscopic media, that is, $m|U_0|\sim\mathcal{O}(1)$ (with the electron mass $m$), is realistic. Then, the second refractive index is not necessarily suppressed for low energies in continuous media, but must be considered on an equal footing with the remaining modes. This behavior will become more transparent for the purely spacelike case to be investigated below.

\subsubsection{\label{section-time-like-case-propagation-modes}Propagation modes}

In order to examine the propagation modes for the purely timelike sector, we can employ \eqref{time3} in the matrix of \eqref{iso-matrix-1}, yielding
\begin{align}
\left[M_{ij}\right]&=-\begin{pmatrix}
n_1^2 & n_1n_2 & n_1n_3 \\
n_1n_2 & n_2^2 & n_2n_3 \\
n_1n_3 & n_2n_3 & n_3^2 \\
\end{pmatrix} \notag \\
&\phantom{{}={}}+2\mu\omega {U}_0(n^2-1)\begin{pmatrix}
\pm n & \mathrm{i}n_3 & -\mathrm{i}n_2 \\
-\mathrm{i}n_3 & \pm n & \mathrm{i}n_1 \\
\mathrm{i}n_2 & -\mathrm{i}n_1 & \pm n \\
\end{pmatrix}\,.
\label{timelike1-4}
\end{align}
Solving $M_{ij}E^{j}=0$, one finds
\begin{subequations}
\label{timelike1-7}
\begin{align}
E_{y}&=\frac{\pm\mathrm{i} n_{3} n - n_{1}n_{2}}{n^{2}-n_{1}^{2}} E_{x}\,, \\[1ex]
E_{z}&=\frac{\mp\mathrm{i} n_{2} n - n_{1}n_{3}}{n^{2}-n_{1}^{2}} E_{x}\,,
\end{align}
\end{subequations}
such that the normalized electric fields ${\bf{E}}_{\pm}$ of the propagating waves are given by
\begin{equation}
{\bf{E}}_{\pm}=\frac{1}{\sqrt{2}n\sqrt{n^2-n_1^2}}\begin{pmatrix}
n^{2}-n_{1}^{2} \\
\pm\mathrm{i}n_{3}n-n_1n_2 \\
\mp\mathrm{i}n_{2}n-n_1n_3 \\
\end{pmatrix}\,.  \label{field-105B}
\end{equation}
The latter coincide exactly with those of \eqref{field-105} except of the labels being switched. Basically, for the timelike configuration, the electric-field modes of the MCFJ and MCFJ-type higher-derivative electrodynamics are the same, despite the different refractive indices of these theories. Note that $\mathbf{E}_{\pm}$ of Eqs.~(\ref{field-105}), (\ref{field-105B}) do not depend on $V_0$ and $U_0$, respectively. The refractive index illustrated in Fig.~\ref{figure-time-like-2} is associated with the electric field $\mathbf{E}_-$ of Eq.~(\ref{field-105B}), whereas those of Figs.~\ref{figure-time-like-1}, \ref{figure-time-like-3} are linked to $\mathbf{E}_+$.

For a direct physical  interpretation of these propagating modes, let us choose again a convenient coordinate system where propagation occurs along the $z$ axis, i.e., let $\mathbf{n}$ be given by \eqref{field-146}. In this system, the normalized electric fields are
\begin{align}
{\bf{E}}_{\pm} &= \frac{1}{\sqrt{2}}\begin{pmatrix}
1 \\
\pm \mathrm{i} \\
0
\end{pmatrix}\,,
\label{field-166B}
\end{align}
which are the same as those stated in~\eqref{field-117}. These are polarization vectors for a left-handed and right-handed circular polarization, respectively, typical of optically active media. Such an optical activity can be expressed in terms of the rotatory power of \eqref{eq:rotatory-power1}, if the refractive indices $n_{+}$ and $n_{-}$ are known. It is worthwhile to note that, although \eqref{time4} provides, in general, three refractive indices, there are only two distinct electric-field configurations, those of \eqref{field-166B}. There are still three propagating modes, one associated with each refractive index. We will come back to this aspect in the forthcoming section, too.

\subsection{\label{section-space-like-case}Purely spacelike case}

Let us now consider the purely spacelike scenario for the background vector, $ {U}_{0} = 0$ and ${\bf{U}}\neq {\bf{0}}$, and also $\tilde{\epsilon} \mapsto \epsilon$ (setting $\sigma=0$). Then \eqref{iso-dispersion-relation-1} yields
\begin{align}
0&=\epsilon (n^{2}-\mu \epsilon)^{2}- 4 (n^{2}-1)^{2} \mu \omega^{2}  \nonumber \\
&\phantom{{}={}}\times\left[ (\mu \epsilon-n^2) {\bf{U}}^{2} + ({\bf{n}}\cdot {\bf{U}})^{2} \right]\,. \label{spacelike1}
\end{align}
Implementing ${\bf{n}}\cdot {\bf{U}}= n | {\bf{U}}| \cos\theta$ in \eqref{spacelike1}, we obtain
\begin{subequations}
\label{prop-spacelike9A}
\begin{equation}
\label{prop-spacelike9A-11}
n^{2}-\mu\epsilon= \pm 2 \mu \omega (n^{2}-1) |{\bf{U}}| \alpha\,,
\end{equation}
where we have defined
\begin{equation}
\alpha^{2}\equiv 1-\frac{n^{2}}{\mu\epsilon}\sin^{2}\theta\,.
\label{prop-spacelike10A}
\end{equation}
\end{subequations}
With this parameterization, we can straightforwardly analyze two special cases: (i) the perpendicular configuration where ${\bf{n}}\cdot {\bf{U}}=0$ and $\sin^{2}\theta=1$; (ii) the longitudinal configuration with $\sin^{2}\theta=0$ and $\bf{n} \cdot {\bf{U}} = \pm |{\bf{n}}| | {\bf{U}}|$ where the plus (minus) sign holds for ${\bf{n}}$ parallel (antiparallel) to ${\bf{U}}$. These choices can provide some physical insights on the behavior of electromagnetic-wave propagation.

To obtain the propagation modes, we again work in a coordinate system where \eqref{field-146} holds. Then the matrix (\ref{iso-matrix-1}) simplifies as
\begin{align}
[M_{ij}] &= \begin{pmatrix}
n_{3}^{2}-\mu\epsilon && -\mathrm{i} \mu\epsilon \beta(\omega) U_{3} && \mathrm{i} \mu\epsilon \beta(\omega) U_{2} \\
\mathrm{i} \mu\epsilon \beta(\omega) U_{3} && n_{3}^{2}-\mu\epsilon && - \mathrm{i} \mu\epsilon \beta(\omega) U_{1} \\
-\mathrm{i} \mu\epsilon \beta(\omega) U_{2} && \mathrm{i} \mu\epsilon \beta(\omega) U_{1} && - \mu\epsilon
\end{pmatrix}\,, \label{extra-matrix-1}
\end{align}
where $\beta(\omega)=2\omega (n_{3}^{2}-1)/\epsilon$. For each case parameterized with $\alpha$, we can insert \eqref{prop-spacelike9A-11} into \eqref{extra-matrix-1} and solve $M_{ij}E^{j}=0$ to achieve the electric fields of the corresponding modes.

\subsubsection{\label{section-perpendicular-configurations}$\mathbf{U}$-perpendicular configuration}

\begin{figure}
	\begin{centering}
		\includegraphics[scale=0.68]{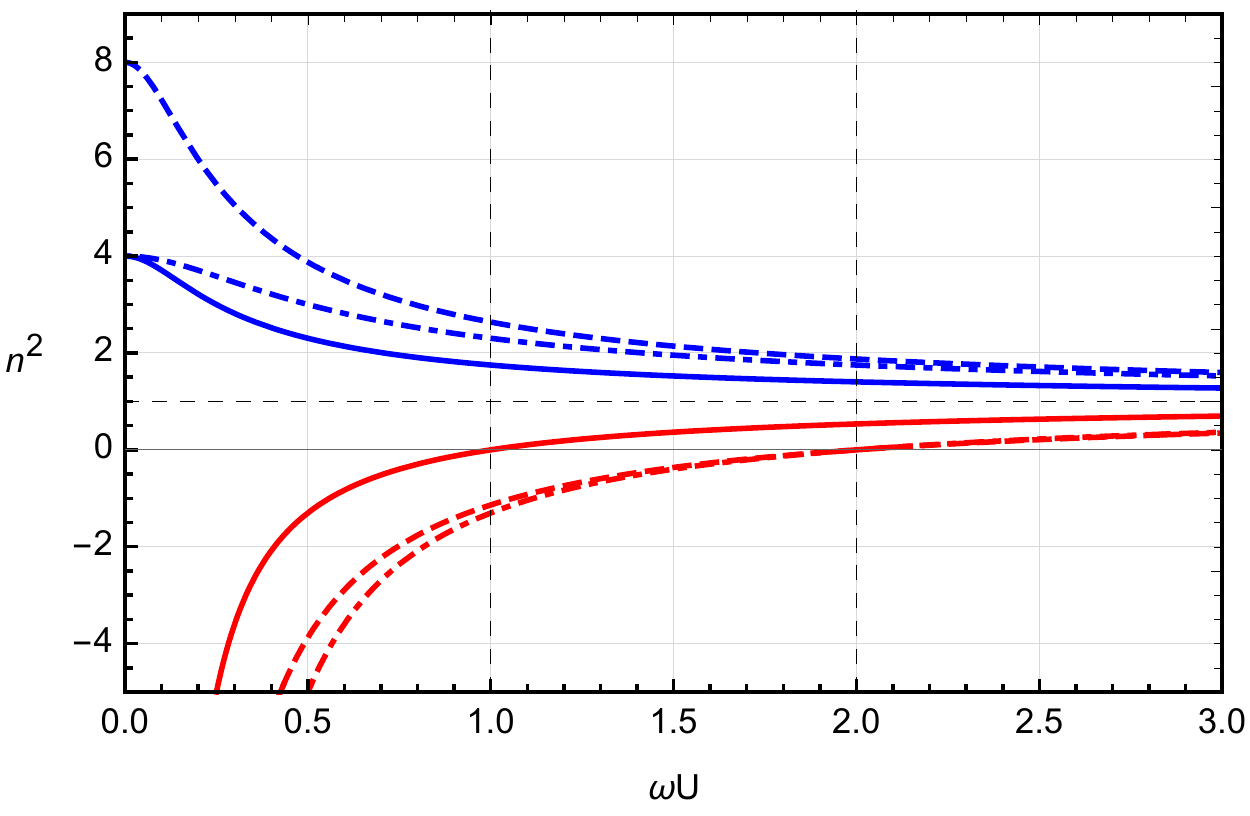}
		\par\end{centering}
	\caption{\label{figure-space-like-perpendicular} Behavior of the refractive indices $n^{2}_{\pm}$ of \eqref{space27} in terms of $\omega U$ where $ U=|{\bf{U}}|$. Blue curves (above the horizontal dashed line) represent $n^{2}_{+}$, while red curves (below the horizontal dashed line) depict $n^{2}_{-}$. For solid lines, $\mu=2$ and $\epsilon=2$; for dashed lines, $\mu=2$ and $\epsilon=4$; for dashed-dotted lines, $\mu=1$ and $\epsilon=4$. Gray vertical dashed lines indicate $\omega_{-} U = 1$ and $\omega_{-}U=2$, respectively, where $\omega_-$ is given by \eqref{space27-1}.}
\end{figure}

First, we consider the orthogonal configuration, i.e., ${\bf{U}}\bot {\bf{n}}$ and $\sin^{2}\theta=1$, so that \eqref{prop-spacelike9A} becomes
\begin{equation}
n^{2}-\mu\epsilon= \pm 2 \mu \omega (n^{2}-1) |{\bf{U}}|\sqrt{1-\frac{n^{2}}{\mu\epsilon}}\,,
\label{prop-spacelike9B}
\end{equation}
which can be written as
\begin{equation}
(n^{2}-\mu\epsilon)\left[ 4 \mu^{2} \omega^{2} {\bf{U}}^{2} (n^2-1)^2 + \mu\epsilon(n^{2} -\mu\epsilon) \right]=0\,, \label{space12-1}
\end{equation}
implying $n^{2}=\mu\epsilon$ and
\begin{align}
0&=4 \mu^{2} \omega^{2} {\bf{U}}^{2} n^{4}+ (\mu\epsilon - 8 \mu^{2} \omega^{2} {\bf{U}}^{2} ) n^{2} \notag \\
&\phantom{{}={}}- \mu^{2}\epsilon^{2} + 4 \mu^{2} \omega^{2} {\bf{U}}^{2}\,.  \label{space20}
\end{align}
The first solution, $n^{2}=\mu\epsilon$, corresponds to the ordinary refractive index of Maxwell electrodynamics in macroscopic media that we denote as $n_0=\sqrt{\mu\epsilon}$. On the other hand, \eqref{space20} captures information stemming from the higher-derivative term of \eqref{higher-7} and the background ${\bf{U}}$, leading to the following solutions:
\begin{subequations}
\label{space27}
\begin{equation}
n^{2}_{\pm} = 1+f_{\pm}\,,
\end{equation}
where
\begin{align}
\label{space27B}
f_{\pm}&=\frac{\epsilon}{8\mu\omega^{2} {\bf{U}}^{2}} \left(-1 \pm \sqrt{1+\Upsilon}\, \right)\,, \\[1ex]
\label{eq:definition-Upsilon}
\Upsilon&=16\mu^{2}\omega^{2} {\bf{U}}^{2}\left(1-\frac{1}{\mu\epsilon}\right)\,.
\end{align}
\end{subequations}
The behavior of $n_{\pm}^{2}$ in terms of the dimensionless parameter $\omega |{\bf{U}}|$ is presented in Fig.~\ref{figure-space-like-perpendicular}. We notice that $n_+$ is real in the entire frequency domain and exhibits anomalous dispersion. Furthermore, the function $n^{2}_{-}$ has a simple root,
\begin{equation}
\label{space27-1}
\omega_{-}= \frac{\epsilon}{2| {\bf{U}}|}\,.
\end{equation}
The latter is interpreted as a critical value (cf.~\eqref{field-170-1-1}), below which $n_{-}$ is purely imaginary, whereupon no propagation occurs. Above $\omega_{-}$, the refractive index $n_{-}$ becomes real. As a consequence, electromagnetic waves can propagate in this regime.

The first vertical dashed line, located at the value $\omega_{-} |{\bf{U}}|=1$, separates the absorption and propagation zones for the mode represented by the solid red line. The second vertical dashed line, in $\omega_{-} |{\bf{U}}|=2$, does so for the modes depicted by the dashed and dashed-dotted lines. In detail, we observe that:

\begin{itemize}

\item For $0<\omega < \omega_{-}$: the refractive index $n_{+}$ is real and $n_{-}$ is purely imaginary; thus, only the mode associated with $n_{+}$ propagates in this range.
\item For $\omega > \omega_{-}$:  one has $n_{\pm}^{2}>0$ and both modes propagate.
\item In the limit of very low frequencies, ${\omega |{\bf{U}}|\mapsto 0}$, it holds that $n_{+}=\sqrt{\mu\epsilon}$, recovering the usual refractive index of a simple continuous medium in standard electrodynamics.
\item In the limit of very high frequencies, $\omega|\mathbf{U}|\mapsto\infty$, the behavior of the refractive indices is
\begin{equation}
n_{\pm}=1\pm \frac{1}{4\omega|\mathbf{U}|}\sqrt{\epsilon\left(\epsilon-\frac{1}{\mu}\right)}\,.
\label{Uinfty}
\end{equation}
Unsurprisingly, the high-frequency behavior of \eqref{Uinfty}, $n_{\pm}\mapsto 1$, differs from that of the refractive indices of MCFJ theory of Eq.~(\ref{field-170B}) in macroscopic media, which is given by $n_{\pm}\mapsto \sqrt{\mu\epsilon}$. Thus, the impact of a nontrivial permeability and permittivity is suppressed in this regime of the MCFJ-type theory in \eqref{lagrangian3} endowed with higher-derivative operators.

\end{itemize}

With regards to the propagating modes, \eqref{prop-spacelike10A} yields
\begin{align}
\alpha_{\pm}=\sqrt{1-\frac{n_{\pm}^{2}}{\mu\epsilon}}=\sqrt{1-\frac{1+f_{\pm}}{\mu\epsilon}}\,,
\end{align}
indicating different values of $\alpha$ for the distinct refractive indices $n_{\pm}$ of \eqref{space27}. Taking $\mathbf{n}$ as given in \eqref{field-146}, the background has the form ${\bf{U}}=(U_{1},U_{2},0)$ such that ${\bf{n}}\cdot {\bf{U}}=0$. The following propagation modes are then achieved:
\begin{align}
\label{modes-perpendicular-higher-derivative1}
{\bf{E}}_{\pm}&= E_{0}' \begin{pmatrix}
U_2 \\
-U_{1} \\
-2\mathrm{i}\omega f_{\pm}(U_{1}^{2}+U_{2}^{2})/\epsilon
\end{pmatrix} \notag \\
&=\tilde{E}_0'\left(\hat{\mathbf{U}}\times\hat{\mathbf{n}}-2\mathrm{i}\omega f_{\pm}\frac{|\mathbf{U}|}{\epsilon}\hat{\mathbf{n}}\right)\,,
\end{align}
with $f_{\pm}$ given by \eqref{space27B} and the unit vector $\hat{\mathbf{U}}$ pointing along the direction of $\mathbf{U}$. In case we choose the background vector of the simple form ${\bf{U}}=(0,U_{2}, 0)$, \eqref{modes-perpendicular-higher-derivative1} provides
\begin{equation}
\label{modes-perpendicular-higher-derivative1B}
{\bf{E}}_{\pm} = E_{0}' \begin{pmatrix}
1 \\
0  \\
-2\mathrm{i}\omega f_{\pm}U_{2}/\epsilon
\end{pmatrix}\,.
\end{equation}
The latter correspond to transverse, linear polarization modes with additional longitudinal components, in analogy to the mode $\mathbf{E}_-$ of \eqref{field-158-1-3B}. Now, by comparing \eqref{modes-perpendicular-higher-derivative1} to
\eqref{field-158-1-4} obtained for the MCFJ theory in macroscopic matter, we spot intriguing similarities. Our interpretation is that the single mode of \eqref{field-158-1-4} splits into the two of \eqref{modes-perpendicular-higher-derivative1} as a result of the higher-derivative nature of this theory. To understand these modes better, it is reasonable to perform Taylor expansions for $\mathbf{U}\mapsto \mathbf{0}$. Investigating the behavior of $f_{\pm}$ in \eqref{space27B} provides
\begin{equation}
f_+\simeq \epsilon\mu-1\,,\quad f_-\simeq -\frac{\epsilon}{4\mu\omega^2\mathbf{U}^2}+1-\epsilon\mu\,,
\end{equation}
giving rise to
\begin{equation}
n_{+}\mapsto \sqrt{\mu\epsilon}\,,\quad n_{-}\mapsto \sqrt{2-\frac{\epsilon}{4\mu\omega^2\mathbf{U}^2}-\epsilon\mu}\,.
\label{n+n-limit}
\end{equation}
As a consequence, the mode described by $\mathbf{E}_+$ has a well-defined limit for $\mathbf{U}\mapsto 0$, whereas the second mode associated with $\mathbf{E}_-$ does not. Here it is also evident that $n_-$ becomes complex in this regime. So such as for the purely timelike sector, we again encounter a mode whose counterpart \textit{in vacuo} would frequently be denoted as spurious. The situation is different in macroscopic matter, though, because $m|\mathbf{U}|\sim\mathcal{O}(1)$ can be realistic. As before, the second mode must be interpreted as a regular, propagating mode.

Finally, we discuss the first solution $n_0=\sqrt{\mu\epsilon}$ of \eqref{space12-1}. In this case, \eqref{prop-spacelike10A} provides $\alpha=0$. Hence, $M_{ij}E^{j}=0$ implies
\begin{align}
\begin{pmatrix}
0 && 0 && \mathrm{i} U_{2} \\
0 && 0 && - \mathrm{i} U_{1} \\
-\mathrm{i} U_{2} && \mathrm{i} U_{1} && -1/\beta
\end{pmatrix}
\begin{pmatrix}
E_{x} \\
E_{y} \\
E_{z} \\
\end{pmatrix} = 0\,,
\label{extra-mode-1}
\end{align}
yielding the following propagating mode:
\begin{align}
{\bf{E}}_{0} &=\frac{1}{|\mathbf{U}|}\begin{pmatrix}
U_1 \\
U_{2} \\
0
\end{pmatrix}=\hat{\mathbf{U}}\,.
\label{extra-mode-2}
\end{align}
The latter is a linearly polarized mode related to the refractive index $n_0=\sqrt{\mu\epsilon}$ and it is perpendicular to the propagation direction of \eqref{field-146}. Also, one finds ${\bf{E}}_{0} \cdot {\bf{E}}_{\pm}^{*}=0$, with ${\bf{E}}_{\pm}$ given by \eqref{modes-perpendicular-higher-derivative1}. The mode of \eqref{extra-mode-2} is equivalent to that of \eqref{field-158-1-3} found for MCFJ theory in macroscopic matter. Thus, this particular mode remains unaffected by the presence of the additional derivatives in the CFJ-type field operator of \eqref{higher-7}. Also, even in the limit $\mathbf{U}\mapsto 0$, the electric fields $\mathbf{E}_0,\mathbf{E}_{\pm}$ are still governed by the direction $\hat{\mathbf{U}}$. However, $\hat{\mathbf{U}}$ does then not indicate a preferred direction, anymore. Instead, the components $U_1,U_2$ take the role of parameterizing the plane orthogonal to the propagation direction~$\hat{\mathbf{n}}$.

In total, the number of the physical modes in the MCFJ-type theory defined by \eqref{lagrangian3} amounts to~3. Two of these approach the behavior of a standard isotropic medium in the limit $\mathbf{U}\mapsto \mathbf{0}$. In particular, it is the mode associated with $n_{+}$ in \eqref{n+n-limit} and that linked to $n_0=\sqrt{\mu\epsilon}$ of \eqref{space12-1}. Having three propagating modes does not indicate a breakdown of gauge invariance of the theory defined by \eqref{lagrangian3}. The operator of \eqref{higher-7} is clearly gauge-invariant. The third mode originates from the presence of the d'Alembertian in \eqref{higher-7} increasing the polynomial order of the dispersion equation. \textit{In vacuo}, the third mode could be denoted as spurious, but this technical term is misleading in macroscopic matter where the coefficients $m\mathbf{U}$ can take values of $\mathcal{O}(1)$.

As the associated modes are not circularly polarized, birefringence for this case is better characterized in terms of the phase shift per unit length given by \eqref{phase-shift1} instead of the rotatory power in \eqref{eq:rotatory-power1}. We introduce
\begin{equation}
\frac{\Delta_{a,b}}{d}\equiv \frac{2\pi}{\lambda_0}(n_a-n_b)\,,
\end{equation}
where $a,b\in \{0,+,-\}$. Since there are three propagating modes for $\omega>\omega_-$, we can define the following phase shifts acquired after propagation (divided by the propagation distance $d$):
\begin{subequations}
\label{birefringence-2-1}
\begin{align}
\frac{\Delta_{\pm,0}}{d}&=\frac{2\pi}{\lambda_0}\left[\sqrt{1+f_{\pm}}-\sqrt{\mu\epsilon}\right]\,, \\[2ex]
\frac{\Delta_{+,-}}{d}&= \frac{2\pi}{\lambda_0} \left[ \sqrt{ 1 + f_{+} } -\sqrt{ 1 + f_{-} } \right]\,,
\end{align}
\end{subequations}
which are valid in the range where $n_{-}$ is real, i.e., $\omega |{\bf{U}}| > \epsilon /2$. In the limit of high frequencies, $(\omega | {\bf{U}}|)^{-1} \ll 1$,  \eqref{birefringence-2-1} yields
\begin{subequations}
\label{rotatory-space-like-perpendicular-1}
\begin{align}
\frac{\Delta_{\pm,0}}{d}&=\frac{2\pi}{\lambda_0}(1-\sqrt{\mu\epsilon})\pm \frac{\Delta}{2d}\,, \\[2ex]
\frac{\Delta_{+,-}}{d}&=\frac{\Delta}{d}\,,
\end{align}
with
\begin{equation}
\frac{\Delta}{d}\equiv \frac{\pi}{\lambda_0 \omega |{\bf{U}}| } \sqrt{\epsilon \left(\epsilon - \frac{1}{\mu} \right)}\,.
\end{equation}
\end{subequations}
Comparing the modes labeled with $\pm$ to the standard mode, there is a zeroth-order contribution that only involves the permittivity and permeability of the medium.

For $\omega < \omega_-$ (or $\omega | {\bf{U}}| < \epsilon/2$), $n_{-}$ is purely imaginary. Then from \eqref{space27}, $n_{-}$ is rewritten as
\begin{equation}
n_{-}=\mathrm{i}\sqrt{-1-f_{-}}\,.
\label{dichroism-space-like-perpendicular-2}
\end{equation}
Since $\mathrm{Im}(n_{+})=0$ for the full frequency domain, only the mode labeled with the minus sign undergoes attenuation, which is quantified by the absorption coefficient, $\gamma = 2\omega \mathrm{Im}(n_{-})$, that is
\begin{equation}
\gamma = 2\omega \sqrt{ \frac{\epsilon}{8\mu \omega^{2} {\bf{U}}^{2}} \left(1+ \sqrt{1+ \Upsilon}\right) -1}\,, \label{dichroism-space-like-perpendicular-3}
\end{equation}
with $\Upsilon$ given by \eqref{eq:definition-Upsilon}.
In the limit of low frequencies, $\omega | {\bf{U}}|\ll  1$, \eqref{dichroism-space-like-perpendicular-3} can be expanded as
\begin{equation}
\gamma \simeq \frac{1}{| {\bf{U}}|}\sqrt{\frac{\epsilon}{\mu}}\left[ 1 + 2\left(1-\frac{2}{\mu\epsilon}\right)\omega^{2} \mu^{2} | {\bf{U}}|^{2} \right]\,. \label{dichroism-space-like-perpendicular-4}
\end{equation}
It is important to note that the absorption coefficient of \eqref{dichroism-space-like-perpendicular-4} is evaluated in the limit  $\omega |{\bf{U}}|\ll 1$, while the phase shift in \eqref{rotatory-space-like-perpendicular-1} is determined in the opposite limit $(\omega | {\bf{U}}|)^{-1}\ll 1$. Attenuation takes place for a purely imaginary $n_{-}$ and birefringence occurs when $n_{-}$ is real. The condition $\omega | {\bf{U}}| = \epsilon/2$ states a clear cutoff separating the frequency regimes for each effect from each other.

\subsubsection{$\mathbf{U}$-longitudinal configuration}

\begin{figure}[t]
\begin{centering}
\includegraphics[scale=0.68]{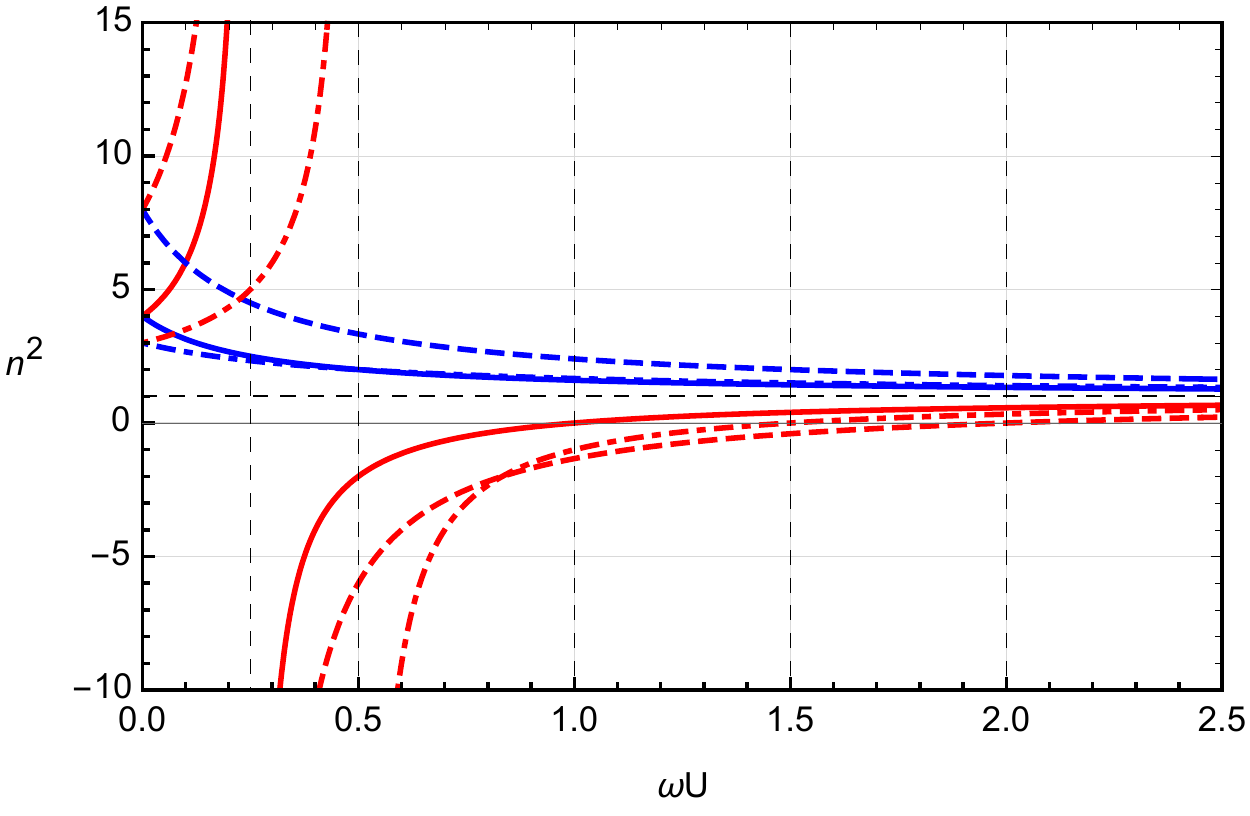}
\par\end{centering}
\caption{\label{figure-space-like-longitudinal} Plot of $n^{2}_{\pm}$ of \eqref{space61} in terms of $\omega U$ with $U=| {\bf{U}}|$. The blue curves, which are positive and monotonically decreasing for the entire frequency range, represent $n_{+}^{2}$. The red lines, constituted by positive upper and negative lower branches, illustrate $n^{2}_{-}$. Solid lines: $\mu=2$ and $\epsilon=2$; dashed lines: $\mu=2$ and $\epsilon=4$; dashed-dotted lines; $\mu=1$ and $\epsilon=3$. Gray vertical dashed lines indicate $\omega {U}\in \{1/4,1/2,1,3/2,2\}$.}
 \end{figure}

Let us now consider the configurations where $\sin\theta=0$, i.e., ${\bf{n}}$ and ${\bf{U}}$ are parallel or antiparallel, for which \eqref{spacelike1} is equivalent to
\begin{align}
0&=(1-4\mu^{2} \omega^{2} {\bf{U}}^{2} )n^4- 2 ( \mu \epsilon - 4 \mu^{2} \omega^{2} {\bf{U}}^{2} ) n^{2} \notag \\
&\phantom{{}={}}+ \mu^{2}\epsilon^{2}-4 \mu^{2} \omega^{2} {\bf{U}}^{2}\,,
\label{space-longitudinal-1}
\end{align}
whose solutions for $n^{2}$ are
\begin{equation}
n_{\pm}^{2} = \frac{\mu(\epsilon \pm 2\omega | {\bf{U}}|)}{ 1 \pm 2\mu\omega | {\bf{U}}| }\,.
\label{space61}
\end{equation}
The behavior of $n^{2}_{\pm}$ in terms of the dimensionless parameter $\omega| {\bf{U}}|$ is displayed in Fig.~\ref{figure-space-like-longitudinal}, for some parameter values.

In this scenario, the mode associated with $n_{+}$ exhibits anomalous dispersion and propagates in the full frequency range, since $n_{+}^{2}>0$.

The mode associated with $n_{-}^{2}$ has two branches. In the superior branch, defined in the frequency range $0<\omega<\omega_{0}$, the mode propagates, with $n^{2}_{-}$ increasing very rapidly with $\omega$. Here,
\begin{equation}
\omega_{0} = \frac{1}{2\mu | {\bf{U}}|}\,, \label{space27-2}
\end{equation}
is the value for which $n^{2}_{-}$ diverges. In Fig.~\ref{figure-space-like-longitudinal}, the first vertical dashed line, given by $\omega_{0} | {\bf{U}}|=1/4$, is asymptotic to the red solid as well as the red dashed curve where the associated functions have singularities at this point and change their signs. The second vertical dashed line is in $\omega_{0} | {\bf{U}}|=1/2$, being asymptotic to both the red upper and lower dashed-dotted curves.
When $\omega>\omega_{0}$ one has $n^{2}_{-}<0$ whose lower branch becomes a purely imaginary refractive index $n_{-}$, representing a nonpropagating mode. This behavior is characteristic in the range $\omega_{0} < \omega < \omega_{-}$, with
\begin{equation}
\omega_{-}=\frac{\epsilon}{2 | {\bf{U}}|}\,, \label{space27-3}
\end{equation}
being the root of \eqref{space61}. Equation~(\ref{space27-3}) stands for the cutoff frequency above which the mode associated with $n_{-}$ propagates. The third and fourth vertical dashed lines, given by $\omega_{-}|{\bf{U}}|=1$ and $\omega_{-}|{\bf{U}}|=1.5$, indicate the beginning of the propagation regime for the red solid curve and the red dashed curve, respectively.

Regarding the $\mathbf{U}$-longitudinal propagation modes, for which ${\bf{n}}\cdot {\bf{U}}= \pm n | {\bf{U}}|$, one takes $\alpha=1$ as well as ${\bf{U}}=(0,0,{U}_{3})$ for $\mathbf{n}$ given by \eqref{field-146}. In this case, the resulting modes are
\begin{equation}
\mathbf{E}_{\pm}=\frac{1}{\sqrt{2}}\begin{pmatrix}
1\\
\mp \mathrm{i} \\
0
\end{pmatrix}\,, \label{prop-spacelike35-5}
\end{equation}
representing right-handed and left-handed circularly polarized waves, respectively. Hence, when ${\bf{U}}$ and ${\bf{n}}$ point along the same direction, the modes become transverse again such that their polarizations are perpendicular to~${\bf{n}}$.

Now, in order to describe birefringence effects, we evaluate the rotatory power by inserting \eqref{space61} into \eqref{eq:rotatory-power1}, that is,
\begin{subequations}
\label{birefringence-3-1}
\begin{equation}
\delta= - \frac{\sqrt{\mu\epsilon}}{2}\omega \left[ \frac{g_{+}}{{\sqrt{1+2\mu\omega |{\bf{U}}|}}} -\frac{g_{-}}{{\sqrt{1-2\mu\omega |{\bf{U}}|}}} \right]\,,
\end{equation}
where
\begin{equation}
g_{\pm} = \sqrt{1 \pm \frac{2\omega}{\epsilon} | {\bf{U}}|}\,. \label{birefringence-3-1B}
\end{equation}
\end{subequations}
The latter result holds for the regions where $n_{-}$ is real, that is, for $\omega < \omega_{0}$ and $\omega > \omega_{-}$, according to Fig.~\ref{figure-space-like-longitudinal}. In the limit $\omega | {\bf{U}}|\ll 1$, \eqref{birefringence-3-1} provides a rotatory power nonlinear in the frequency, namely:
\begin{equation}
\delta\simeq\sqrt{\frac{\mu}{\epsilon}}(\mu\epsilon-1)\omega^2|{\bf{U}}|\,.
\label{expansion-rotatory-longitudinal-1}
\end{equation}
As already mentioned, the refractive index $n_{-}$ is purely imaginary in the range $\omega_{0} < \omega < \omega_{-}$, constituting an absorption zone, which is explicitly given by
\begin{align}
 \frac{1}{2\mu| {\bf{U}}|} < \omega < \frac{\epsilon}{2| {\bf{U}}|}\,. \label{dichroism-space-like-longitudinal-1}
\end{align}
In this regime the refractive index reads
\begin{align}
n_{-} &= \mathrm{i} \sqrt{\mu\epsilon} \sqrt{\frac{1-2\omega |{\bf{U}}|/\epsilon}{2\mu\omega |{\bf{U}}|-1}}\,,\label{dichroism-space-like-longitudinal-4}
\end{align}
and the corresponding dichroism coefficient is
\begin{align}
\delta_{\mathrm{d}} = \frac{\sqrt{\mu\epsilon}}{2} \omega \sqrt{\frac{1 - 2\omega | {\bf{U}}|/\epsilon}{2\mu\omega | {\bf{U}}|-1}}\,. \label{dichroism-space-like-longitudinal-5}
\end{align}

\section{\label{final-remarks}Final Remarks}

In this work, we examined an electrodynamics of continuous media based on Maxwell equations modified by \textit{CPT}-odd terms, whereas the usual constitutive relations ${\bf{D}}=\epsilon {\bf{E}}$ and ${\bf{H}}=\mu^{-1} {\bf{B}}$ were assumed to hold. At first, we reviewed some basic properties of the MCFJ model, followed by an analysis of the dimension-five higher-derivative extension of MCFJ electrodynamics. Our general focus was on describing electromagnetic-wave propagation in matter governed by these \textit{CPT}-odd modifications.

In Sec.~\ref{section-comparison-to-MCFJ-1}, we examined MCFJ electrodynamics, given by the Lagrangian of Eq.~(\ref{MCFJ-1}), in a continuous medium with a fixed background $V^{\mu}=(V^{0}, {\bf{V}})$ present. To analyze the propagation behavior of electromagnetic waves, we obtained the dispersion relations and the refractive indices for two scenarios: (i) a timelike background, $V^{\mu}=(V^{0}, {\bf{0}})$ and (ii) a spacelike background, $V^{\mu}=(0, {\bf{V}})$.  For scenario (i), the refractive indices are always real, giving rise to propagation without losses as well as birefringence. The corresponding rotatory power, $\delta=-\mu V_{0}/2$, is frequency-independent. For scenario (ii), one refractive index, $n_{+}$, is always real, while the other, $n_{-}$, may be complex, corresponding to an absorption regime. In this case, birefringence and dichroism occur in different frequency ranges. The rotatory power and dichroism coefficient are both frequency-dependent.

In Sec.~\ref{section4}, we considered an electrodynamics in a ponderable medium modified by a MCFJ-type higher-derivative term of dimension five, given by the Lagrangian of Eq.~(\ref{lagrangian3}). After writing up the altered Maxwell equations, a sixth-order dispersion equation was achieved.	
In the purely timelike scenario, ${U}^{\mu}=({U}^{0}, {\bf{0}})$, studied in Sec.~\ref{section-time-like-case}, we obtained a third-order equation in the refractive index $n$ providing three solutions. One solution is real for any frequency, while the remaining two are complex for some frequency range (absorption range). This behavior occurs even for a dielectric nonconducting substrate.
Such an effect is represented, for example, by Figs.~(\ref{figure-time-like-1}) and (\ref{figure-time-like-3}), where the graphs indicate that $\mathrm{Im}[n(\omega)] \neq 0$ in the absorption range $\omega_{-} < \omega < \omega_{+}$, with $\omega_{\pm}$ given in \eqref{time14}. That property is entirely ascribed to the higher-derivative coupling term, since the usual MCFJ electrodynamics in ponderable media does not exhibit an absorption regime for a purely timelike background. Comparing Figs.~(\ref{plot-MCFJ-timelike}) and (\ref{figure-time-like-all-together}) with each other allows us to notice the differences between the propagating modes in the usual and higher-derivative timelike case. Furthermore, the propagation modes obtained correspond to left-handed and right-handed circular polarizations [see Sec. \ref{section-time-like-case-propagation-modes}].

In Sec.~\ref{section-space-like-case}, we addressed the purely spacelike scenario, governed by an involved dispersion relation. It was analyzed for two particular cases: (a) the perpendicular configuration, where ${\bf{n}} \cdot {\bf{U}}=0$, and (b) the longitudinal configurations, ${\bf{n}}\cdot {\bf{U}}=\pm n |{\bf{U}}|$. In scenario (a), one finds $n_{+}^{2} >0$ for all frequencies, which indicates the absence of absorption for this mode. On the other hand, $n_{-}$ becomes purely imaginary for $\omega<\omega_{-}$, with $\omega_{-}$ defined in \eqref{space27-1}. Absorption occurs in this range for the mode associated with $n_{-}$ (see Fig.~\ref{figure-space-like-perpendicular}). Hence, attenuation and birefringence are expected in the regions $\omega<\omega_{-}$ and $\omega > \omega_{-}$, respectively. For scenario (b), the refractive index $n_{+}$ is always real, as well, while $n_{-}$ exhibits two distinct branches separated by the frequency $\omega_{0}$ given in \eqref{space27-2}. The upper branch, defined for $\omega < \omega_{0}$, is characterized by a region of sharp normal dispersion. In $\omega = \omega_{0}$, the refractive index $n_{-}$ diverges. In the lower branch, the mode associated with $n_{-}$ turns complex and returns to the propagation regime for $\omega > \omega_{-}$. This mode possesses different physical behaviors (propagation or absorption). Birefringence occurs for $\omega < \omega_{0}$ and $\omega >\omega_{-}$, while absorption takes place for $\omega_{0} < \omega < \omega_{-}$. In both ranges, the acquired phaseshift between different modes and the absorption coefficient are frequency-dependent.

In order to compare the spacelike configurations of the dimension-three and five MCFJ electrodynamics, we examined Figs.~\ref{plot-MCFJ-spacelike-perpendicular} and \ref{figure-space-like-perpendicular}. The dimension-three model shows normal dispersion, while in the dimension-five framework modes emerge that exhibit both anomalous and normal dispersion. The absorption zones are qualitatively analogous to each other in both cases. Comparing Figs.~\ref{plot-MCFJ-spacelike-longitudinal} and \ref{figure-space-like-longitudinal}, we notice that dimension-three and five modes are characterized by normal and anomalous dispersion, while only the higher-derivative model exhibits two branches of normal dispersion. In the limit of high frequencies, one has $n_{\pm}^{2} \mapsto \mu\epsilon$ based on \eqref{eq:refractive-index} for dimension-three MCFJ electrodynamics and $n_{\pm}^{2} \mapsto 1$ inferred from Eqs.~(\ref{space27}), (\ref{space61}) for dimension-five MCFJ-type electrodynamics. These findings allow us to distinguish between the two models. Therefore, the presence of higher derivatives implies a richer plethora of frequency-dependent propagating modes.

\section*{Acknowledgments}

The authors thank the anonymous referee for helpful comments that contributed to improving the paper and clarifying some results. The authors also express their gratitude to FAPEMA, CNPq and CAPES (Brazilian research agencies) for invaluable financial support. In particular, M.M.F. is supported by FAPEMA Universal 01187/18 and CNPq Produtividade 311220/2019-3. M.S. appreciates support by FAPEMA Universal 00830/19 and CNPq Produtividade 312201/2018-4. Furthermore, the authors are indebted to CAPES/Finance Code 001.

\appendix

\section{\label{AppendixA}Covariant Maxwell equations in matter}

Here we derive the Maxwell equations and the constitutive relations from Eqs.~(\ref{e1}) and (\ref{e2}).
Equation~(\ref{e2}) implies
\begin{subequations}
\begin{align}
G^{0i}&=\frac{1}{2} \chi^{0i\alpha\beta}F_{\alpha\beta}\,, \label{a1} \\
G^{0i}&= \frac{1}{2}\chi^{0i0j}F_{0j}+\frac{1}{2}\chi^{0ij0}F_{j0}+\frac{1}{2}\chi^{0imn}F_{mn}\,, \label{a2}
\end{align}
\end{subequations}
which can be simplified by using the symmetry properties of the tensor  $\chi^{\mu\nu\varrho\sigma}$, i.e., \eqref{eq:symmetry-property-chi-2}. We also implement
\begin{equation}
\label{eq:definitions-electric-magnetic-field}
F_{0i}=-F_{i0}=E^{i}\,,\quad F_{mn}=-\epsilon_{mnk}B^{k}\,,
\end{equation}
where $\epsilon_{mnk}$ is the three-dimensional Levi-Civita symbol. Thus, \eqref{a2} becomes
\begin{subequations}
\begin{align}
G^{0i}&=-\chi^{0ij0}E^{j}-\frac{1}{2} \chi^{0imn} \epsilon_{mnk}B^{k}\,, \label{a3}\\
G^{0i}&=- D^{i}\,, \label{a5}
\end{align}
where we have defined the electric displacement field $D^{i}$, which involves the medium's response to applied electromagnetic fields, as
\begin{equation}
\label{a6}
D^{i}=\chi^{0ij0} E^{j}+\frac{1}{2}\chi^{0imn} \epsilon_{mnk}B^{k}\,.
\end{equation}
\end{subequations}
From \eqref{a6} we can define the electric permittivity $\epsilon_{ij}$ as well as the tensor $\gamma_{ij}$ describing the magnetic contribution to electric displacement field (see \eqref{constitutive1}) as
\begin{equation}
\label{a7}
\epsilon_{ij}\equiv \chi^{0ij0}, \quad \gamma_{ik} \equiv \frac{\chi^{0imn}\epsilon_{mnk}}{2}\,.
\end{equation}
The antisymmetric nature of $\chi^{\mu\nu\alpha\beta}$ allows us to write
\begin{equation}
G^{\mu\nu}=-G^{\nu\mu}\,. \label{a10}
\end{equation}
Now we can evaluate the components $G^{ij}$. In doing so, we get
\begin{subequations}
\begin{align}
G^{ij}&=\frac{1}{2} \chi^{ij\alpha\beta}F_{\alpha\beta}\,, \label{a11} \\
G^{ij}&=\frac{1}{2} \chi^{ij0k}F_{0k}+\frac{1}{2} \chi^{ijk0}F_{k0}+\frac{1}{2}\chi^{ijmn}F_{mn}\,, \label{a12}
\end{align}
\end{subequations}
which is recast by using \eqref{eq:symmetry-property-chi-2} as well as \eqref{eq:definitions-electric-magnetic-field}. Then,
\begin{equation}
G^{ij}=-\chi^{ijk0}E^{k}-\frac{1}{2}\chi^{ijmn}\epsilon_{mnk}B^{k}\,.
\label{a13}
\end{equation}
In order to obtain a relation between $G_{ij}$ and $H^{i}$ similar to that between $F_{ij}$ and $B^{i}$, let us now contract \eqref{a13} with $\epsilon_{ijl}$ such that
\begin{equation}
\epsilon_{ijl}G^{ij}=-\frac{2}{2}\epsilon_{ijl}\chi^{ijk0}E^{k}-\frac{2}{4}\epsilon_{ijl}\chi^{ijmn}\epsilon_{mnk}B^{k}\,, \label{a15}
\end{equation}
where we introduced the factor $(2/2)$ in each term of \eqref{a15}. The motivation for doing so will become clear shortly, as this manipulation allows us to write down an expression very similar to $F_{mn}=-\epsilon_{mnk}B^{k}$, but for the components $G_{ij}$ and $H^{i}$. Thus, we define the magnetic permeability $\mu_{ij}$ as well as $\tilde{\gamma}_{ij}$ governing the electric contribution to the magnetic field (see \eqref{constitutive2}) as
\begin{align}
(\mu^{-1})_{lk}\equiv \frac{1}{4} \epsilon_{ijl}\chi^{ijmn}\epsilon_{mnk}, \quad \tilde{\gamma}_{lk}\equiv \frac{\epsilon_{ijl}\chi^{ijk0}}{2}\,. \label{a16}
\end{align}
Then \eqref{a15} simplifies as
\begin{subequations}
\begin{align}
\epsilon_{ijl}G^{ij}&=-2(\mu^{-1})_{lk}B^{k}-2\tilde{\gamma}_{lk}E^{k}\,, \label{a17} \\[1ex]
\epsilon_{ijl}G^{ij}&=-2 H^{l}\,, \label{a18}
\end{align}
where we have defined the magnetic field $H^{l}$, which describes the medium's response to applied electromagnetic fields via
\begin{equation}
H^{l}=(\mu^{-1})_{lk}B^{k}+\tilde{\gamma}_{lk}E^{k}\,. \label{a19}
\end{equation}
\end{subequations}
Let us contract \eqref{a18} with $\epsilon_{lmn}$, whereupon
\begin{subequations}
\begin{align}
\epsilon_{lmn}\epsilon_{ijl}G^{ij}&=-2 \epsilon_{lmn}H^{l}\,, \label{a20} \\[1ex]
G^{mn}&=-\epsilon_{mnl}H^{l}\,, \label{a23}
\end{align}
\end{subequations}
where we have used \eqref{a10}.

Now that we have expressed  the constitutive relations in terms of the constitutive tensor $\chi^{\mu\nu\alpha\beta}$, we can derive the field equations associated with the Lagrange density of \eqref{e1}. Thus, we start by rewriting \eqref{e1}:
\begin{align}
\mathcal{L}&=-\frac{1}{8} \chi^{\mu\nu\alpha\beta}F_{\alpha\beta}F_{\mu\nu}-A_{\mu}J^{\mu}, \notag \\
&=-\frac{1}{8} \chi^{\mu\nu\alpha\beta} \partial_{\alpha}A_{\beta}\partial_{\mu}A_{\nu}+\frac{1}{8} \chi^{\mu\nu\alpha\beta} \partial_{\alpha}A_{\beta}\partial_{\nu}A_{\mu} \nonumber \\
&\phantom{{}={}}+\frac{1}{8} \chi^{\mu\nu\alpha\beta} \partial_{\beta}A_{\alpha}\partial_{\mu}A_{\nu}-\frac{1}{8} \chi^{\mu\nu\alpha\beta} \partial_{\beta}A_{\alpha}\partial_{\nu}A_{\mu} \notag \\
&\phantom{{}={}}-A_{\mu}J^{\mu}\,. \label{a26}
\end{align}
We rename the indices ($\nu \leftrightarrow \mu$) in the second and fourth term of \eqref{a26} and after that we employ the symmetry property of \eqref{eq:symmetry-property-chi-1}. This gives us
\begin{equation}
\mathcal{L}= -\frac{1}{4} \chi^{\mu\nu\alpha\beta} \partial_{\alpha}A_{\beta}\partial_{\mu}A_{\nu}+\frac{1}{4}\chi^{\mu\nu\alpha\beta} \partial_{\beta}A_{\alpha}\partial_{\mu}A_{\nu}-A_{\mu}J^{\mu}\,, \label{a27}
\end{equation}
which can be simplified by replacing $\alpha \leftrightarrow \beta$ in the second term and using \eqref{eq:symmetry-property-chi-2}. Hence, we finally obtain
\begin{align}
\mathcal{L}&=-\frac{1}{2}\chi^{\mu\nu\alpha\beta}\partial_{\alpha}A_{\beta}\partial_{\mu}A_{\nu}-A_{\mu}J^{\mu}\,. \label{a28}
\end{align}
Using the Euler-Lagrange equations
\begin{equation}
\frac{\partial \mathcal{L}}{\partial A_{\kappa}}-\partial_{\rho} \left( \frac{\partial \mathcal{L}}{\partial (\partial_{\rho} A_{\kappa})} \right)=0\,, \label{a29}
\end{equation}
one arrives at
\begin{align}
\frac{\partial \mathcal{L}}{\partial (\partial_{\rho} A_{\kappa})}&=-\frac{1}{2} \left( \chi^{\beta\alpha\rho\kappa} \partial_{\beta}A_{\alpha}+\chi^{\rho\kappa\alpha\beta}\partial_{\alpha}A_{\beta} \right)\,, \label{a32}
\end{align}
where we have relabeled $\mu \rightarrow \beta$, $\nu \rightarrow \alpha$ in the first term on the right-hand side. Now, we also implement \eqref{eq:symmetry-property-chi-3} in the first contribution, and in the second term we take advantage of \eqref{eq:symmetry-property-chi-2}. Therefore, \eqref{a32} provides
\begin{equation}
\frac{\partial \mathcal{L}}{\partial (\partial_{\rho} A_{\kappa})}=-\frac{1}{2} \chi^{\rho\kappa\beta\alpha} F_{\beta\alpha}=-G^{\rho\kappa}\,, \label{a34}
\end{equation}
and one also finds
\begin{equation}
\frac{ \partial \mathcal{L}}{\partial A_{\kappa}}=-J^{\mu} \delta_{\mu\kappa}=-J^{\kappa}\,. \label{a35}
\end{equation}
So using Eqs.~(\ref{a34}) and (\ref{a35}) in \eqref{a29}, we finally get the covariant form of the Maxwell equations in simple matter:
\begin{equation}
\partial_{\rho}G^{\rho\kappa}=J^{\kappa}\,. \label{a36}
\end{equation}
Taking $\kappa=0$, one finds Gauss's law:
\begin{subequations}
\begin{align}
\partial_{i}G^{i0}&=J^{0}\,, \label{a38} \\[1ex]
\nabla \cdot {\bf{D}}&=\rho\,, \label{a40}
\end{align}
\end{subequations}
where we have employed \eqref{a5} and $J^{\mu}=(\rho, {\bf{J}})$. Amp\`ere's law is obtained by taking $\kappa = i$ in \eqref{a36}, that is
\begin{subequations}
\begin{align}
\partial_{0} G^{0i}+\partial_{j} G^{ji}&=J^{i}\,, \label{a42} \\[1ex]
\partial_{t} (-D^{i} ) - \partial_{j} (\epsilon_{jik} H^{k} )&= J^{i}\,, \label{a43}
\end{align}
\end{subequations}
where we have used Eqs.~(\ref{a5}), (\ref{a10}), and (\ref{a23}). Applying further simplifications to \eqref{a43}, yields
\begin{subequations}
\begin{align}
\epsilon_{ijk}\partial_{j}H^{k}-\partial_{t}D^{i}&=J^{i}\,, \label{a44} \\[1ex]
\nabla \times {\bf{H}} - \partial_{t}{\bf{D}}&= {\bf{J}}\,. \label{a45}
\end{align}
\end{subequations}

\section{\label{AppendixB} Rotatory power and dichroism coefficient}

As mentioned at the end of Sec.~\ref{section2}, when the propagating modes resulting from an electromagnetic theory are left-handed and right-handed circularly polarized waves, birefringence is characterized in terms of the rotatory power while absorption is described via the dichroism coefficient, presented in Eqs.~(\ref{eq:rotatory-power1}) and (\ref{eq:dichroism-power1}), respectively. Such relations can be derived by means of the polarization vectors of a wave traveling through a medium. Consider, for instance, a linearly polarized wave propagating through a medium along the $z$ axis. Hence, the initial electric field can be written as
\begin{subequations}
\begin{equation}
{\bf{E}}_{i}= {\bf{E}}_{0i} \mathrm{e}^{\mathrm{i} (kz-\omega t)}\,, \label{rotation-1}
\end{equation}
with the polarization vector (for an electric field pointing along the $x$ axis):
\begin{align}
{\bf{E}}_{0i}=\begin{pmatrix}
1\\
0\\
0
\end{pmatrix} = \frac{1}{2} \begin{pmatrix}
1\\
-\mathrm{i} \\
0
\end{pmatrix} + \frac{1}{2} \begin{pmatrix}
1 \\
\mathrm{i} \\
0
\end{pmatrix}\,, \label{rotation-2}
\end{align}
\end{subequations}
which corresponds to the sum of polarization vectors associated with left-handed and right-handed circular polarizations, respectively. After the wave passes through a distance $z$ in the medium, the final electric field is a linear combination of two components, ${\bf{E}}_{+}$ and ${\bf{E}}_{-}$, with the wave vectors ${\bf{k}}_{+}$ and ${\bf{k}}_{-}$, respectively. One then has
\begin{align}
{\bf{E}}_{f} &= {\bf{E}}_{+} \mathrm{e}^{\mathrm{i}( k_{+} z - \omega t)}+ {\bf{E}}_{-} \mathrm{e}^{\mathrm{i} (k_{-} z - \omega t)}, \notag \\
{\bf{E}}_{f}&=\frac{1}{2} \begin{pmatrix}
1\\
\mathrm{i} \\
0
\end{pmatrix} \mathrm{e}^{\mathrm{i} k_{+} z} \mathrm{e}^{-\mathrm{i}\omega t} + \frac{1}{2} \begin{pmatrix}
1\\
-\mathrm{i}\\
0
\end{pmatrix} \mathrm{e}^{\mathrm{i}k_{-} z} \mathrm{e}^{-\mathrm{i}\omega t}\,,  \label{rotation-3}
\end{align}
which can be cast into the form
\begin{subequations}
\begin{align}
{\bf{E}}_{f} &= \frac{1}{2} \mathrm{e}^{\mathrm{i} \psi} \mathrm{e}^{-\mathrm{i}\omega t} \left[  \mathrm{e}^{-\mathrm{i} \theta} \begin{pmatrix}
1 \\
\mathrm{i} \\
0
\end{pmatrix}  +  \mathrm{e}^{\mathrm{i} \theta} \begin{pmatrix}
1\\
-\mathrm{i} \\
0
\end{pmatrix}\right], \notag \\
{\bf{E}}_{f}&=\mathrm{e}^{\mathrm{i}\psi} \mathrm{e}^{-\mathrm{i}\omega t} \begin{pmatrix}
\cos \theta \\
\sin \theta \\
0
\end{pmatrix}\,,  \label{rotation-10}
\end{align}
with the quantities
\begin{align}
\theta &\equiv -\frac{( k_{+}-k_{-})z}{2}\,,  \label{rotation-5} \displaybreak[0]\\
\psi &\equiv \frac{(k_{+}+k_{-})z}{2}\,.  \label{rotation-6}
\end{align}
\end{subequations}
Notice that \eqref{rotation-10} describes a linearly polarized wave whose polarization vector is rotated by an angle $\theta$. From \eqref{rotation-5}, one obtains
\begin{equation}
\theta=- \frac{ (n_{+}-n_{-}) z \omega}{2}\,, \label{rotation-11}
\end{equation}
where we have used ${\bf{k}}=\omega {\bf{n}}$. In general, the refractive indices can be complex quantities. Because of this, one can infer from \eqref{rotation-11}
\begin{equation}
\frac{\theta}{z}= -\frac{\omega}{2} \left[\mathrm{Re} (n_{+}) + \mathrm{i}  \mathrm{Im}(n_{+}) - \mathrm{Re} (n_{-}) - \mathrm{i}  \mathrm{Im}(n_{-})\right]\,, \label{rotation-12}
\end{equation}
from which we define the specific rotatory power stated in \eqref{eq:rotatory-power1} as well as the dichroism coefficient of \eqref{eq:dichroism-power1}.
Notice that when the medium is nonbirefringent, $\theta=0$ and $\psi=k z$. Then, the form of \eqref{rotation-1} is recovered from \eqref{rotation-10}.

\end{document}